\documentclass[10pt,journal,compsoc]{IEEEtran}
\ifCLASSOPTIONcompsoc
\usepackage[nocompress]{cite}
\else
\usepackage{cite}
\fi
\ifCLASSINFOpdf
\usepackage[pdftex]{graphicx}
\graphicspath{{./}}
\DeclareGraphicsExtensions{.pdf,.jpeg,.png}
\else
\fi
\usepackage{amsmath}
\usepackage{algorithmic}

\usepackage{array}
\usepackage{url}

\usepackage{amsfonts}
\usepackage{amssymb}
\usepackage{amsthm}
\usepackage{enumerate}
\usepackage{float}
\usepackage{color}
\usepackage{stfloats,fancyhdr}
\usepackage{multirow}
\usepackage{changepage}
\usepackage[normalem]{ulem}
\usepackage{subcaption}

\usepackage{url}
\usepackage{hyperref}

\newcommand{\txtColor}{black}
\newcommand{\FigPath}{./}
\newcommand{\figWidth}{0.95\linewidth}
\newcommand{\oneFigWidth}{0.8\linewidth}
\newcommand{\twoColumnFigWidth}{0.95\textwidth}
\newcommand{\structFigWidth}{0.95\linewidth}

\newcommand{\autoTrig}{Auto-trigger }

\begin{document}
	%
	\title{ROFA: An OFDMA system for Ultra-Reliable Wireless Industrial Networking}
	%
	%
	%
	%
	
	\author{Jiaxin~Liang,~
			and~Soung~Chang~Liew,~\IEEEmembership{Fellow,~IEEE}
		\IEEEcompsocitemizethanks{\IEEEcompsocthanksitem Jiaxin Liang, and Soung Chang Liew are with Department of Information Engineering, The Chinese University of Hong Kong, Hong Kong SAR, China \protect\\
			E-mail: \{jiaxin, soung\}@ie.cuhk.edu.hk
			\IEEEcompsocthanksitem Copyright (c) 20xx IEEE. Personal use of this material is permitted. However, permission to use this material for any other purposes must be obtained from the IEEE by sending a request to pubs-permissions@ieee.org.}
		}
	
	%
	%

	\markboth{Journal of \LaTeX\ Class Files,~Vol.~14, No.~8, August~2015}%
	{Shell \MakeLowercase{\textit{et al.}}: Bare Demo of IEEEtran.cls for Computer Society Journals}
	%



	\IEEEtitleabstractindextext{%
		\begin{abstract}
			This paper proposes and demonstrates a PHY-layer design of a real-time prototype that supports Ultra-Reliable Communication (URC) in wireless infrastructure networks. The design makes use of Orthogonal Frequency Division Multiple Access (OFDMA) as a means to achieve URC. Compared with Time-Division Multiple Access (TDMA), OFDMA concentrates the transmit power to a narrower bandwidth, resulting in higher effective SNR. Compared with Frequency-Division Multiple Access (FDMA), OFDMA has higher spectrum efficiency thanks to the smaller subcarrier spacing. Although OFDMA has been introduced in 802.11ax, the purpose was to add flexibility in spectrum usage. Our \textbf{R}eliable \textbf{OF}DM\textbf{A} design, referred to as \textit{ROFA}, is a clean-slate design with a single goal of ultra-reliable packet delivery. ROFA solves a number of key challenges to ensure the ultra-reliability: (1) a downlink-coordinated time-synchronization mechanism to synchronize uplink transmission of users, with at most $0.1us$ timing offset; \textcolor{\txtColor}{(2) an ``STF-free'' packet reception synchronization method that makes use of the property of synchronous systems to avoid packet misdetection;} and (3) an uplink precoding mechanism to reduce the CFOs between users and the AP to a negligible level. We implemented ROFA on the Universal Software Radio Peripheral (USRP) SDR platform with real-time signal processing. \textcolor{\txtColor}{Extensive experimental results show that ROFA can achieve ultra-reliable packet delivery ($PER<10^5$) with $11.5dB$ less transmit power compared with OFDM-TDMA when they use $3$ and $52$ subcarriers respectively.}
		\end{abstract}
		
		\begin{IEEEkeywords}
			ultra-reliable communication, industrial wireless system, OFDMA, software-defined radio.
		\end{IEEEkeywords}}

	\maketitle

	\IEEEdisplaynontitleabstractindextext

	%
	\IEEEpeerreviewmaketitle

	\IEEEraisesectionheading{\section{Introduction}\label{sec:introduction}}

	%
	%
	%
	%
	\IEEEPARstart{U}{ltra-Reliable} Communication (URC), first proposed in \cite{popovski_ultra-reliable_2014}, has gained much attention since its introduction. It is a key enabler for critical message delivery in applications that demand highly reliable communication \cite{saad_vision_2020}, such as mission-critical industrial applications. These applications may involve the delivery of sensor measurements and process states to a monitoring and control center \cite{brown_ultra_reliable_2017}, and it is of paramount importance that this information can reach the control center with minimum failure rates. To date, such mission-critical communication is often supported by wired networks. For futuristic applications that demand portability and mobility, however, wireless networking is the only option. The design of URC wireless networks, therefore, is of much interest. 
	
	This paper studies the physical-layer (PHY) design of URC wireless networks. We believe that Orthogonal Frequency Division Multiple Access (OFDMA) is a good option for URC wireless networking, and we address several key issues related to the PHY-layer design of reliable OFDMA wireless networks. 

	OFDMA is a multiple access scheme inspired by OFDM. In OFDMA, the OFDM spectrum is shared by multiple users, with each carrying its signal on a subset of the OFDM subcarriers. Compared with Time-Division Multiple Access (TDMA), OFDMA can be more reliable. This is because, for a given overall bandwidth, each user in an OFDMA system uses a smaller chunk of the overall bandwidth to communicate. By concentrating the transmit signal power to this ``narrower bandwidth'', the effective SNR can be higher than that of a TDMA system in which users take turns to use the whole bandwidth. Frequency-Division Multiple Access (FDMA) is an alternative for reliable narrowband communication.  It is, however, less spectrum-efficient than OFDMA.  This is because the subcarrier spacing in OFDMA is smaller, thanks to the use of DFT/IDFT signal processing. 

	Although OFDMA shares many signal processing techniques with Orthogonal Frequency-Division Multiplexing (OFDM), it is not trivial to build a practical OFDMA system because of other challenges not existing in OFDM systems. In this paper, we consider infrastructure networks in which multiple users communicate with an Access Point (AP) through an uplink (UL)---from users to AP---and a downlink (DL)---from AP to users. We assume the UL and DL use the same band, and they take turns using the band. That is, the UL and DL are interspersed in a TDMA manner. For the UL, however, the users' signals overlap in time and are multiplexed using the OFDMA technique. 

	\subsection{Reliable OFDMA Challenges} \label{subsec:challenges}
	The design of OFDMA DL is similar to that of OFDM given that the DL signal is from one single source. The OFDMA UL signal, however, consists of signals from multiple users. As a result, there are a number of distinct OFDMA challenges for the UL, particularly for URC:
	\begin{itemize}
		\item The signals of the UL transmitters (users) in the OFDMA system must be aligned to within-cyclic prefix (CP) \cite{mckinley_2004_evm}. This is not difficult to do if the users use a common clock source to synchronize their transmissions. However, for many indoor industrial applications, particularly those that demand wireless networking, the users are not co-located and they operate based on their own clock driven by a local oscillator. 
		\item Conventional packet detection in OFDM systems (e.g., that used in the receiver of Wi-Fi) applies autocorrelation signal processing on a time-domain Short Training Field (STF) in the packet preamble to detect the the arrival of the packet. However, conventional packet detection is intended for receivers that do not know \textit{a priori} when the packet is coming. The receiver may misdetect the arrival of packets if the channel or the transmitter is in poor condition.
		\item There may be relative Carrier Frequency Offsets (CFO) among the UL signals from multiple users given that the users do not use a common clock/oscillator. Large CFOs among them may cause their subcarriers to be non-orthogonal, giving rise to mutual interferences that compromise the reliability of the system. In addition, the CFO between a user and the AP may also degrade the reliability. 
	\end{itemize}

	\subsection{Solutions to Challenges and Their Validation} \label{subsec:solution_to_challenges}
	To address the above challenges, we design and implement an OFDMA system, named ROFA (\textbf{R}eliable \textbf{OF}DM\textbf{A}). A summary of our contributions and results is as follows:
	\begin{itemize}
		\item To tackle challenge 1, we put forth a downlink-coordinated time-synchronization method to synchronize UL transmissions of users. In this method, the users estimate the arrival time of the DL packet to obtain the starting time of the DL. If the separation of the DL time and UL time is fixed and known, the users can then derive the starting time of the UL based on the starting time of the DL. In this way, the users can synchronize their UL transmissions. Experiment results indicate that our implementation can align the UL packets from multiple users to not just within-CP, but within one sample in a 10MHz system (i.e., to within $0.1 us$).
		\item To address challenge 2, we eliminate the STF in the UL packet's preamble. Correspondingly, we propose an ``STF-free'' mechanism for the AP, named \textit{Auto-trigger} mechanism, for time-synchronization of  UL packet decoding (i.e., to mark the beginning of a packet for the decoding process that follows). Our experiment shows that $55\%$ of the UL packets' arrival times provided by \autoTrig are $0$-sample-shifted and $100\%$ of the UL packets do not exceed $1$-sample shifting in a $10MHz$ system.
		\item To address challenge 3, we leverage the \textit{CFO reciprocity} between DL transmission and UL transmission and apply the estimated DL CFO to precode the UL signals to eliminate the CFO between users and the AP.  To the extent that the CFO between each user and the AP can be reduced to a negligible level, the CFO between users will also be negligible. The key then is accurate DL CFO estimation. We propose a highly accurate DL CFO estimation method, named \textit{SLP CFO estimation}, that uses a combination of signal processing based on STF, Long Training Field (LTF), and a new packet field called \textit{Post-LTF} (P-LTF). Simulations show that SLP CFO estimation outperforms conventional CFO estimation that use only the STF or that uses the STF and LTF, reducing the residual CFO by at least $50\%$ in the high SNR regime ($>15dB$). We found that conventional CFO estimation methods are not accurate enough for reliable UL CFO precoding. The
		SLP CFO estimation, on the other hand, performs very effective UL CFO precoding to allow us to build a highly reliable OFDMA system.
	\end{itemize}

	Besides theoretical and simulation studies of our solutions, we also implemented the solutions on the Universal Software Radio Peripheral (USRP) SDR platform to validate them in a real deployment setting. Importantly, our implementation of ROFA can run in real-time (i.e., all the signal processing of our solutions run in real-time rather than offline). This real-time capability of ROFA is important to validating our solution 3 above. In particular, solution 3 uses the estimation of DL CFO in order to perform UL CFO precoding. If only offline signal processing is possible, we will need to first collect the DL data and estimate the CFO offline. We cannot investigate the effect of UL CFO precoding that follows ``immediately'' after the DL transmission. The CFO may have changed a great deal by the time we use the CFO measured a long time ago to do UL CFO precoding.

	We benchmarked the reliability performance of ROFA against that of an OFDM-TDMA system (i.e., a TDMA scheme in which the users take turns transmitting using all the data subcarriers of an OFDM system). Overall, we found that ROFA can outperform OFDM-TDMA significantly under various experimental settings.
	
	\section{Related Work}\label{sec:related_work}
	\subsection*{802.11ax}\label{subsec:80211_ax_intro}
	Among the commercial off-the-shelf (COTS) communication systems, the latest commercial Wi-Fi standard, 802.11ax, also includes OFDMA as one of its modulation schemes \cite{khorov_tutorial_2019}. Although both 802.11ax and ROFA use OFDMA, ROFA is very different from 802.11ax because of their different goals.

	The 802.11ax standard, first released at the end of 2016 \cite{bellalta_ieee_2016}, introduced OFDMA as a feature. As the successor of the high-throughput WLAN amendment, throughput was a top priority in 802.11ax. It leverages OFDMA to add flexibility to the use of spectrum resource \cite{bellalta_ieee_2016}, so that the total throughput can be further increased \cite{deng_ieee_2014}. For backward compatibility with legacy 802.11 a/b/g/n/ac, however, 802.11ax incurs much overhead. Given that reliability rather than throughput is the premium in many industrial wireless use cases, a clean-slate design that targets reliability without the bloated overhead of 802.11ax is desirable. ROFA is a clean OFDMA design with the sole objective of ultra-reliable communication.
	
	The target application scenarios of 802.11 (including 802.11ax) and ROFA are different. In general, 802.11 is designed to be capable of supporting user traffic that is bursty in nature. For example, network traffic generated by a computer is unpredictable and bursty. If transmission resources are pre-allocated to the user devices without regard to whether they have traffic to send or not, that will lead to much wastage. 802.11ax OFDMA adopts a polling strategy to poll which user devices have traffic to send to the AP . ROFA, by contrast, is targeted for industrial applications where user devices generate steady streams of traffic in a continuous manner (e.g., see \cite{seno_bandwidth_2017, barbosa_exploiting_2016}). The amount of traffic per unit time generated by a user device is predetermined. Given that, we can pre-allocate a number of subcarriers to each user device in a permanent manner to meet the traffic demand. This removes the overhead incurred by a polling strategy.

	Secondly, 802.11ax requires a user to use at least $26$ adjacent subcarriers (1 Resource Unit (RU)) in its OFDMA transmission, for the reason that the resource allocation can be simplified, and the conventional packet detection and CFO compensation algorithms can be reused \cite{khorov_ieee_2015, hoefel_ieee_2016}. However, the minimum size of RU imposes a limit of the transmission reliability, as explained below. In ROFA, we do not impose a minimum number of subcarriers that must be allocated to a user. For example, if a user's traffic rate is very low, we can allocate just one subcarrier to it. With 802.11ax, on the other hand, this user would have to use $26$ subcarriers to transmit intermittently, because its traffic rate is low (i.e., there is much time during which the user device is silent). For ROFA, this user could just use one subcarrier and keeps transmitting using the single subcarrier. This means that the transmit power can be concentrated on a narrowband of that single subcarrier to achieve better reliability than 802.11ax, for which the transmit power is spread over $26$ subcarriers.
	
	
	\begin{figure}
		\centering {\includegraphics[width=\structFigWidth]{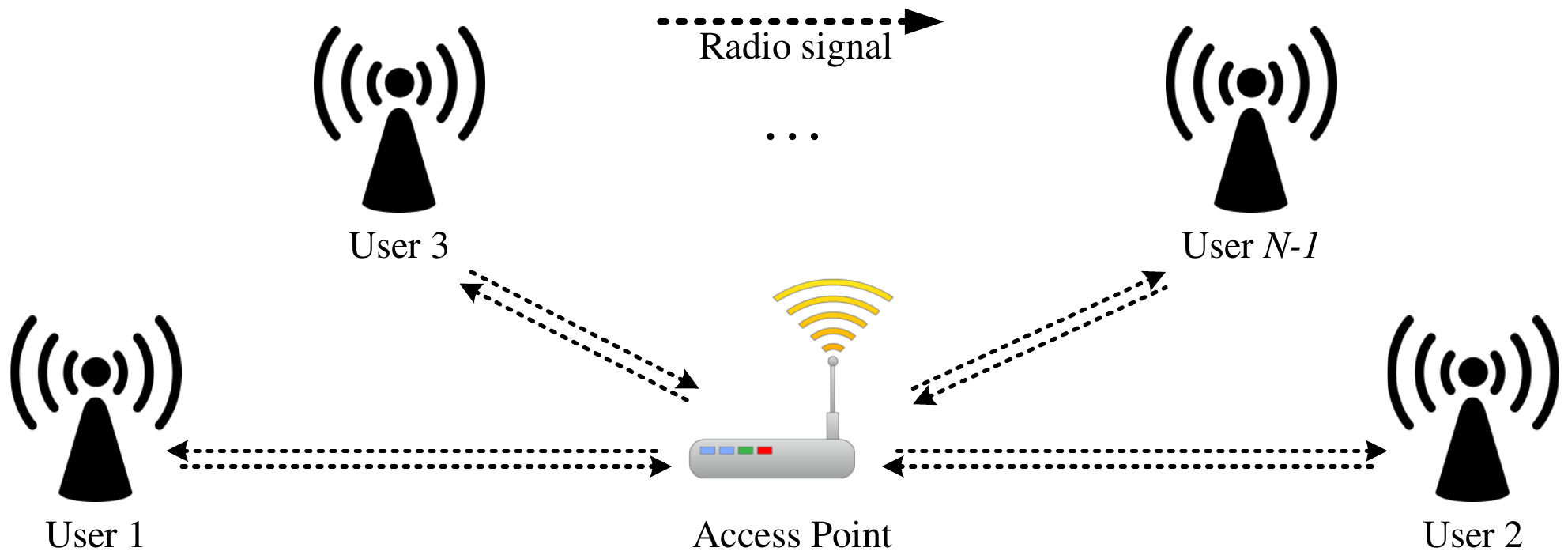}}
		\caption{An example of an n-node infrastructure system. One node is the AP and the other nodes are users.}
		\label{fig:system_arch}
	\end{figure}
	
	\subsection*{Works that Realize Multiuser Wireless Systems on USRP-SDR} \label{subsec:related_sdr_works}
	The past decade has seen development efforts that used the USRP SDR platform to realize multiuser wireless systems. Examples of such systems include Physical-layer Network Coding (PNC) system \cite{you_reliable_2017}, Network-Coded Multiple Access (NCMA) system \cite{you_network-coded_2015}, TDMA system \cite{liang2020design_jiot}, and Non-orthogonal Multiple Access (NOMA) \cite{khan_cooperative_2021, wei_software_2016, reddy_experimental_2021}. The systems in \cite{you_network-coded_2015, you_reliable_2017, liang2020design_jiot, khan_cooperative_2021, wei_software_2016, reddy_experimental_2021}, however, are OFDM-based, not OFDMA-based. Furthermore, their systems were geared toward improving throughput, not reliability.
	
	Recently, the authors in \cite{gokceli_ofdma-based_2016} presented an OFDMA-based network-coded system implemented on USRP. The authors used a common clock wherein the clock signal is distributed over wired links to all the devices. This set-up is for experimental convenience only and is not practical in a distributed system where users are not co-located and are not connected by wires. We remark that the whole idea of using a wireless network is that wiring the user devices is not viable or convenient for the use cases of interest. In ROFA, each user device uses its own local clock source. As a result, the design of ROFA faces a number of challenges that must be overcome (see Section \ref{subsec:challenges}).

The remainder of this paper is organized as follows: Section \ref{sec:system_arch} introduces the system architecture of ROFA. Section \ref{sec:system_design} presents the details of the components that are uniquely designed and implemented for ROFA. Section \ref{sec:reliability_rofa} presents reliability analysis of ROFA. Experimental results are provided in Section \ref{sec:experiment}. Section \ref{sec:conclusion} concludes this paper.

	\section{System Architecture} \label{sec:system_arch}

	\begin{figure}
		\centering {\includegraphics[width=\structFigWidth]{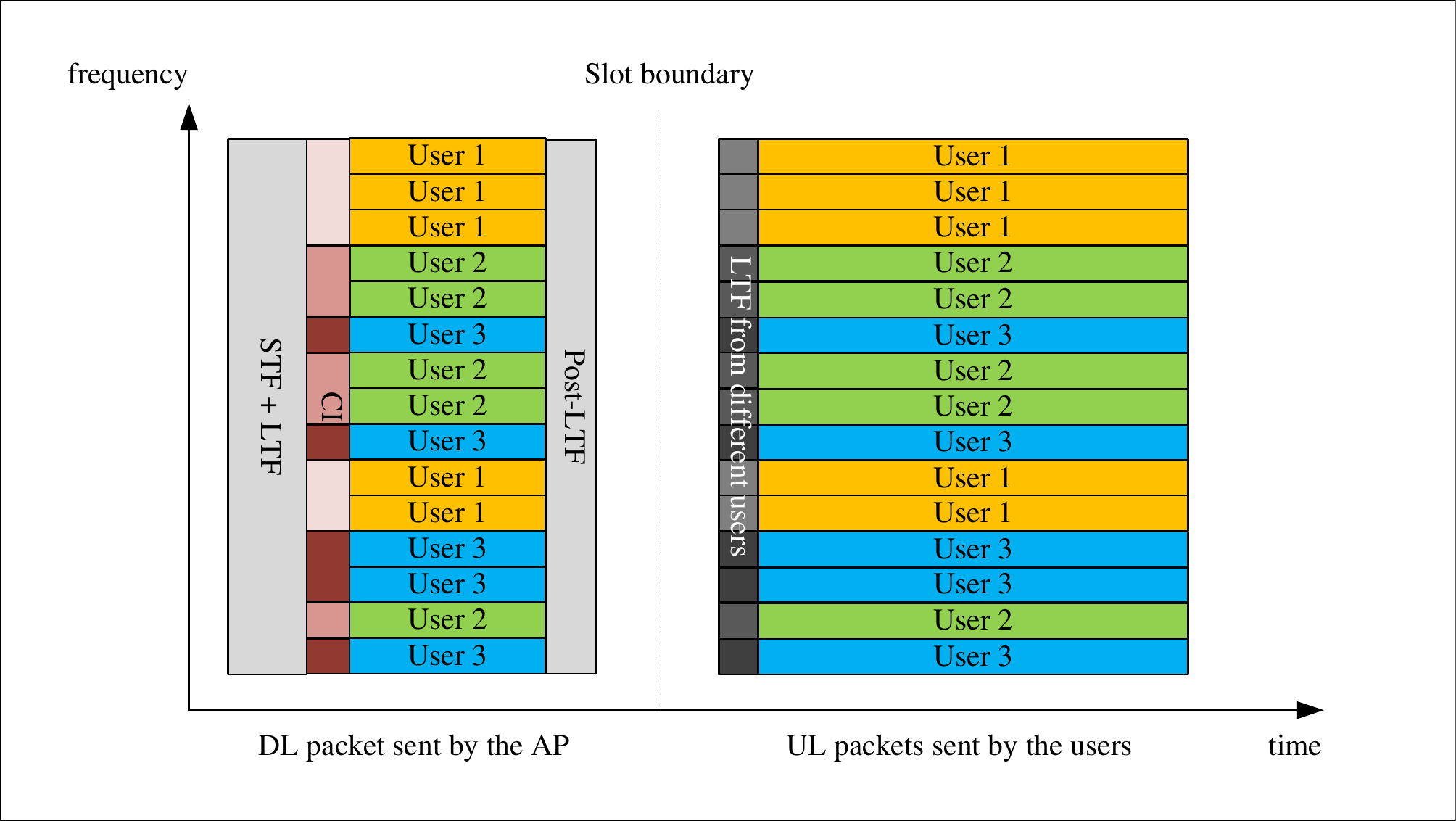}}
		\caption{An example of the DL transmission and UL transmission, where the AP broadcasts a packet to $3$ users and $3$ users use OFDMA to access the UL channel.}
		\label{fig:frame_struct}
	\end{figure}

	Fig. \ref{fig:system_arch} shows the network architecture of our focus throughout this paper. There are $N$ radio nodes in the network, each consisting of one PC and one USRP. One of the radio nodes is the AP and the other radio nodes are users. Let $i$, $i \in \left\{ {1,2,...,N - 1} \right\}$, be the indices of users. The index of the AP is $0$. Let ${N_{{\rm{FFT}}}}$ be the size of FFT, ${N_{{\rm{data}}}}$ be the number of OFDM symbols in the data payload of one packet, and ${\cal S}_i$ be the set containing the indices of subcarriers allocated to user $i$.
	
	In ROFA, the AP and users take turns to access the channel. Specifically, ROFA divides the spectrum resources into time frames. Each frame has two time slots allocated to the \textbf{DL phase} and \textbf{UL phase}, respectively. Fig. \ref{fig:frame_struct} shows an example of a time frame with three users in the network.

	Before diving into the details, we remark that although the major challenges of OFDMA reside in UL, for completeness, we implemented both DL and UL on the USRP SDR platform. Importantly, some techniques for UL transmissions (e.g., UL packets synchronization, UL CFO precoding) rely on channel information retrieved from DL transmissions. The following gives an overview of the design of ROFA for DL followed by that of UL, over the USRP SDR platform.
	
	\subsection{DL Phase} \label{subsec:dl_phase}
	In DL phase, the AP broadcasts a DL packet to all users (see Fig. \ref{fig:frame_struct}). The DL packet uses all the available subcarriers. It carries normal data and control information (CI). It also provides a timing reference to align the transmission of UL packets from the users to the AP. In other words, the DL packet is not only a data packet and a control packet, but is also a beacon packet that provides a timing reference to the users. With OFDMA, the AP can allocate different subcarriers to different users.
	
	The DL packet begins with an STF and an LTF. The STF is for packet detection and the LTF is for channel estimation and packet synchronization (i.e., finding the beginning of a packet). Besides carrying data, the DL packet also piggybacks the control information (CI) (e.g., users' subcarrier allocation). At the end of the DL packet, a Post-LTF (P-LTF) is appended to aid the accurate estimation of the carrier frequency offsets (CFO) between the AP and users. Details about the use of P-LTF for accurate CFO estimation is elaborated in Section \ref{subsubsec:dl_cfo_est}.
	
	\begin{figure*}
		\centering {\includegraphics[width=\twoColumnFigWidth]{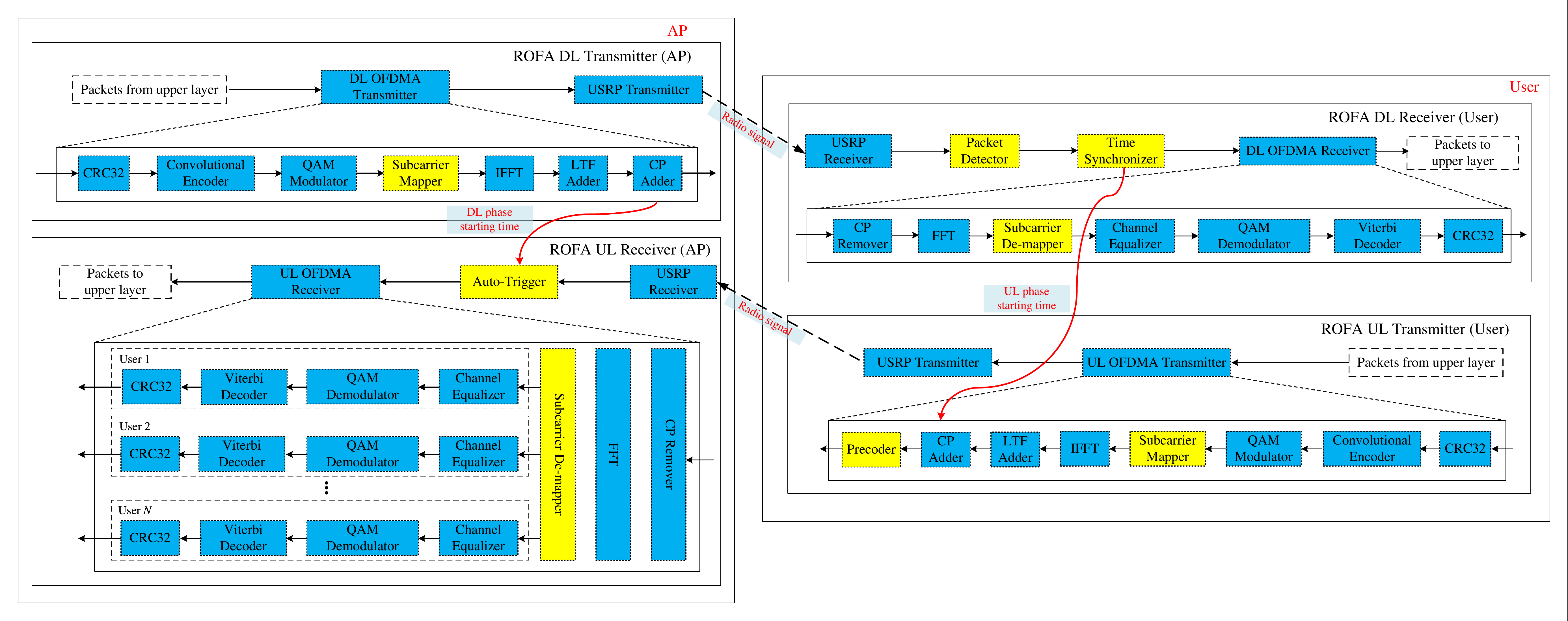}}
		\caption{Block diagram of an AP and a user, which includes the ROFA DL Transmitter/Receiver and the ROFA UL Transmitter/Receiver. The blue blocks contain technologies common to OFDM and OFDMA. The yellow blocks contain technologies unique to ROFA's implementation of OFDMA.}
		 \label{fig:flowgraph}
	\end{figure*}
	
	Fig. \ref{fig:flowgraph} shows the transceiver architecture of ROFA. The upper right part of Fig. \ref{fig:flowgraph} is the block diagram of the ROFA DL receiver. The packet detector detects the arrival of a DL packet in the time domain, followed by a time synchronizer that identifies the beginning of the packet. Recall that the DL packet also serves as a timing reference for the transmission of UL packets. By finding the beginning of the DL packet, the time synchronizer can then extrapolate the beginning time of the upcoming UL phase. Details of the synchronization mechanism can be found in Section \ref{subsec:system_design_ul_sync}.
	
	After the time synchronization, the receiver removes the CP of OFDM symbols and converts the time-domain signal to the frequency-domain signal by FFT. In the subcarrier mapping block, only the symbols on the allocated subcarriers for this user are selected from the output of FFT and passed to the next block. From then on, the user only processes the symbols on the allocated subcarriers to get the packet content.
	
	\subsection{UL Phase} \label{subsec:ul_phase}
	The lower right part of Fig. \ref{fig:flowgraph} is the block diagram of a ROFA UL transmitter. The UL transmitter (user) $i$ obtains a sequence of information bits from its packet source. A CRC32 checksum is computed and appended to the information bits. The information bits sequence are then fed to a convolutional encoder to generate the coded bits. The bit sequence is modulated by a QAM modulator to generate $M$ QAM symbols ${{\bf{b}}_i} = {\left[ {{b_{i,0}},{b_{i,1}},...,{b_{i,M - 1}}} \right]^T}$. After that, the subcarrier mapper divides the QAM symbols ${{\bf{b}}_i}$ into $\left| {{\cal S}_i^{}} \right|$ and feeds the symbol streams to $\left| {{\cal S}_i^{}} \right|$ inputs of the IFFT block that corresponds to subcarriers in the set ${\cal S}_i$; the inputs corresponding to subcarriers not allocated to user $i$ are set to zero.
	
	After the IFFT block, the time-domain OFDM symbols form the payload, and the LTF adder adds an LTF to the beginning of the payload. Note that this LTF is different from the L-LTF for DL. It only occupies the subcarriers allocated to user $i$. Finally, the precoder precodes the time-domain OFDM packet (to compensate for the CFO at the transmitter side, detailed in Section \ref{subsec:system_design_ul_sync}) and sends the precoded packet to the USRP transmitter.
	
	The lower left part of Fig. \ref{fig:flowgraph} is the block diagram of a ROFA UL receiver (AP). The UL receiver (AP) knows when the next UL phase begins by virtue of the fact that the DL transmitter (AP) knows exactly when the preceding DL phase began. The \autoTrig block in Fig. \ref{fig:flowgraph} receives an indicator from the DL transmitter indicating the beginning time of the DL phase (see the red curve in the left part of Fig. \ref{fig:flowgraph}). From the DL-phase beginning time, the DL receiver can compute the next UL-phase beginning time. The \autoTrig block kick-starts the UL reception process when the UL phase begins. (Section \ref{subsubsec:auto_trigger_mech} gives details of the \autoTrig mechanism).
	
	Upon receiving a triggering signal, the UL OFDMA receiver first removes the CP, and then feeds the time-domain samples of the packet to FFT. After the FFT operation, the frequency-domain samples are separated based on the subcarrier allocation scheme (i.e., the set $\left\{ {{\cal S}_1^{},{\cal S}_2^{},...,{\cal S}_{N{\rm{ - }}1}^{}} \right\}$) and fed to $N-1$ parallel processing streams, with each stream containing the data from one user. For each stream, the receiver performs channel equalization, QAM demodulation, Viterbi decoding, and CRC check. The final outcomes of the CRC check are then forwarded to the upper layer.
	
	\section{System Design} \label{sec:system_design}
	In this section, we present the details of the components that are uniquely designed and implemented for ROFA. These components address the challenges mentioned in Section \ref{subsec:challenges}. We first present an UL packet synchronization mechanism for the UL transmitters to align their UL packets (Challenge 1). \textcolor{\txtColor}{Then an UL packet reception synchronization mechanism is provided for the UL receiver to start the packet reception process without STF (Challenge 2).} We then present an UL CFO precoding mechanism to compensate for the UL CFO among the users and the AP (Challenge 3). Finally, we describe the channel estimation and equalization of ROFA UL. 
	
	\subsection{UL Packet Synchronization} \label{subsec:system_design_ul_sync}
	In ROFA, the users use their respective local clock sources and thus have different times. To avoid the inter-block interference (IBI) in the UL transmission of an OFDMA system, the users must ensure that their UL packets can arrive at the AP within the CP duration. In other words, the maximum misalignment of the packet arrival times of different users must be smaller than each other's CP duration.
	
	To ensure the UL packets from different users can meet the time-synchronization requirement, we adapt a time-synchronization mechanism proposed in our previous work, RTTS-SDR \cite{liang2020design_jiot}, for ROFA.
	
	\textit{RTTS-SDR} is a time-slotted system implemented on the USRP SDR platform \cite{liang2020design_jiot}. RTTS-SDR can very accurately align all the users' time slot boundaries at the AP's receiver.  The AP in RTTS-SDR broadcasts beacons periodically as time references and the users adjust their time-slot boundaries according to the arrival times of the beacons. The users then prepare their uplink packets based on the reference time. Implementing time-slot alignment based on beacon arrival times on the USRP SDR platform is non-trivial and involves a number of challenges caused by the uncontrollable delay jitters between the PC and the USRP. Interested readers are referred to \cite{liang2020design_jiot} on the details of these challenges and our solution for them.
	
	A key concept in RTTS-SDR is ``counting before sending''. If a user knows the index of the first sample of the beacon, it can derive the index of the first sample of a future time slot by counting forward a number of samples. For example, if this future time slot is the next UL time slot in ROFA, the user can then position the beginning of its next UL packets with the right sample index.
	
	To adapt the synchronization mechanism of RTTS-SDR for ROFA, the first step is to redefine the number of time slots in a time frame to two. As mentioned in Section \ref{sec:system_arch}, there are only two time slots in a time frame in ROFA, one for DL and one for UL.
	
	The second step is to redefine the packet formats in accordance with the ROFA design. In ROFA, the structure of DL packets and UL packets are different. A DL packet contains L-STF, L-LTF, CI, data, and P-LTF, while an UL packet only contains LTF and data.
	
	Because of the different formats of the DL and UL packets, to improve spectrum efficiency, the durations of the UL slot ${T_{{\rm{UL}}}}$ and the DL slot ${T_{{\rm{DL}}}}$ in ROFA, unlike that in RTTS-SDR, should be different. Thus, the third step is the redefine the time-slot durations of UL and DL.
	
	An added flexibility introduced in ROFA is as follows. The CI in the DL packet contains not only the subcarrier allocation map ${\cal S}_i^{}$, but also the DL and UL packet durations ${T_{{\rm{UL}}}}$ , ${T_{{\rm{DL}}}}$ for the next frame. Upon decoding CI in a DL packet, users can determine their next UL packet's transmission times. In the experiments of this paper, however, we assume that the durations of  ${T_{{\rm{UL}}}}$ , ${T_{{\rm{DL}}}}$ are unchanging. 
	
	\subsection{UL Packet Reception/Decoding Synchronization} \label{subsec:system_design_ul_detect}
	As mentioned in Challenge (2), Introduction, since ROFA targets ultra-reliable packet delivery, misdetecting packets is undesirable. The conventional packet detection mechanism, which detects the arrival of packets based on the received signals, may experience misdetection if the signals are in poor condition. Thus, ROFA uses a more reliable mechanism to synchronize packet decoding with packet arrival. In this subsection, we present our proposed STF-free UL packet synchronization mechanism in ROFA, called the \autoTrig mechanism.
	
	\subsubsection{Eliminating STF} \label{subsubsec:eliminate_stf}
	We could eliminate STF in ROFA UL transmission given that ROFA is a synchronous system. Specifically,\textit{ ROFA is synchronous in which the AP is the provider of the system time. This enables the AP to determine the UL packet arrival time without packet detection.} Doing away with the STS-based detection removes the possibility of misdetection, which could occur once in a while due to noise or other impediments in the UL channel. ROFA does not experience such misdetections.  Furthermore, the removal of STS also reduces packet overhead, allowing more data to pack into a packet for a given packet length.
	
	Conventional packet detection mechanism (based on STF or other methods) is a must  for systems with asynchronous transmissions and receptions, e.g., Wi-Fi systems. In an asynchronous system, the receiver of a packet does not know \textit{a priori} when the packet is coming, and therefore the receiver needs to keep receiving samples from the air and sees if it can detect an incoming packet. The system is asynchronous in the sense that the packets do not arrive synchronously at predictable times. And only when the arrival of a packet is detected does the reception process -- which includes other signal processing such as equalization, FFT, channel decoding -- begin. Thus, a detected packet ``triggers'' the reception process.
	
	ROFA UL, however, is synchronous. Since the AP provides reference times to the users via DL packets and the users align their UL packets based on the reference time, the AP knows the arrival times of UL packets. Furthermore, subcarriers in ROFA UL are pre-allocated to the users on a long-term basis. The AP can go ahead to perform packet decoding without detection. If the CRC does not check out, it could be either due to the error or that the user did not send a packet. Thus, the remaining problem is when to triggers the reception process.
	
	\subsubsection{\textit{Auto-trigger} mechanism} \label{subsubsec:auto_trigger_mech}
	ROFA uses an \autoTrig mechanism. Since the AP determines the starting time of DL transmission, and the durations of DL and UL packets are known, it can compute the arrival time of the next batch of UL packets based on its known information, as elaborated below.
	
	Assuming that the transmission time of the $k$-th DL packet is $t_{{\rm{DL}}}^l$, the AP can compute the arrival time of the $l$-th UL packet $t_{{\rm{UL}}}^k$ by
	\begin{equation}
		t_{{\rm{UL}}}^l = t_{{\rm{DL}}}^k + {T_{{\rm{DL}}}}.
	\end{equation}

	The \autoTrig block at the receiving path obtains $t_{{\rm{DL}}}^k$ from the transmitting path and computes the new future UL packet arrival time $t_{{\rm{UL}}}^l$. ROFA uses the same \textit{sample counting} mechanism as the Sample Counter in \cite{liang2020design_jiot} to get the USRP hardware time of the received samples. In particular, for both the transmit path and the receive path, each sample can be associated with a USRP time (i.e., there is a one-to-one mapping from a sample index to the hardware time). For the transmit path, the hardware time of a sample is the time at which that sample is transmitted by the USRP. For the receive path, the hardware time of a sample is the time at which that sample is received by the USRP.
	
	The \autoTrig block continuously counts the incoming samples from the input. When the time of the incoming sample ${T_b}$ is equal to $t_{{\rm{UL}}}^l$, it generates a peak signal in the control signal path to trigger the execution of the packet reception blocks.
	
	\begin{figure}
		\centering {\includegraphics[width=\structFigWidth]{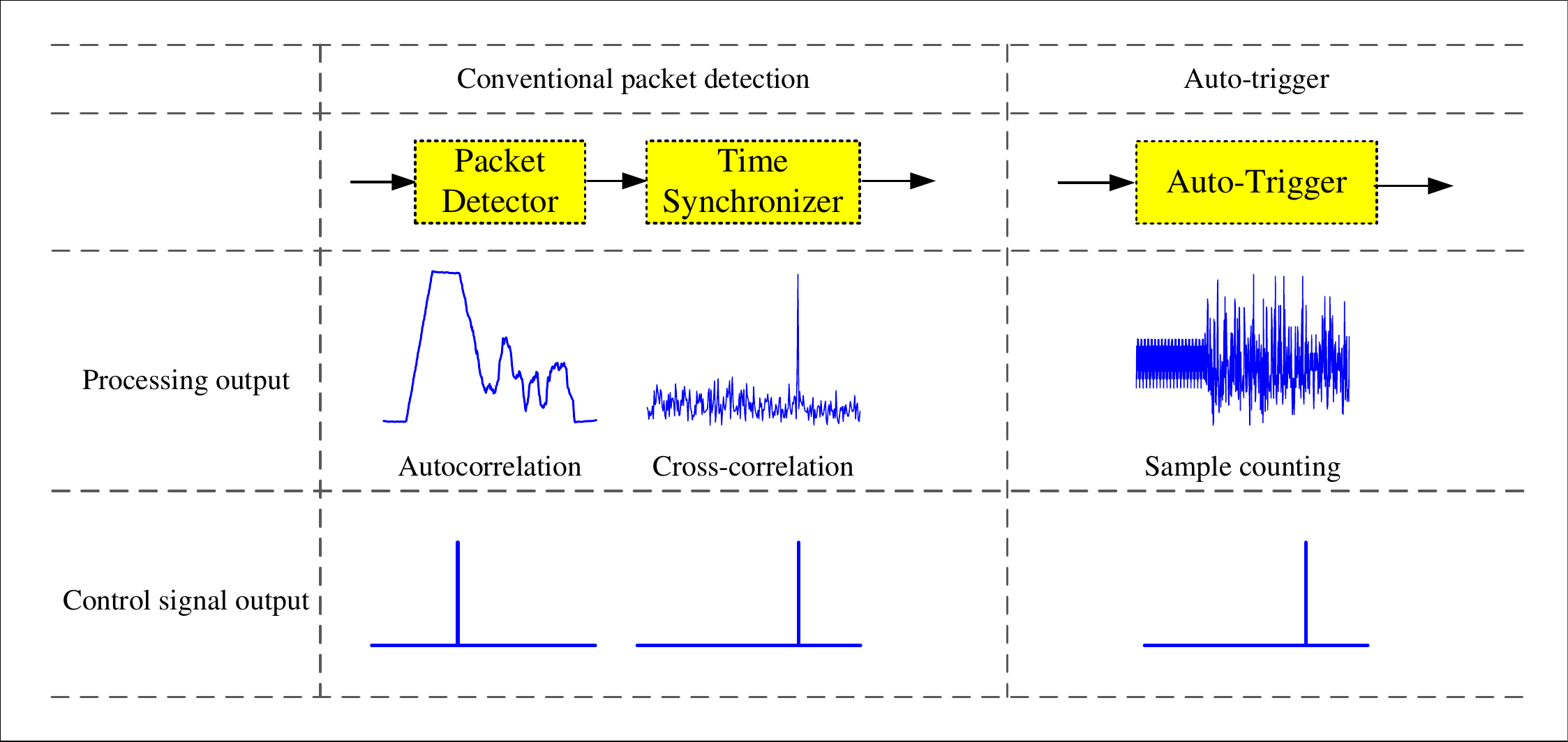}}
		\caption{Comparison between the conventional packet detection and the \autoTrig mechanism.}
		\label{fig:auto_trigger}
	\end{figure}

	Fig. \ref{fig:auto_trigger} compares the packet-reception triggering processes of the convention methods and our \autoTrig method. The control signal output shown in Fig. \ref{fig:auto_trigger} is connected to the subsequent blocks and used to kick-start the subsequent blocks' processing. Specifically, this control signal output needs to keep generating a stream of bits in parallel with the signal sample stream (one control bit corresponding to one data sample) to inform the subsequent blocks of the arrival of a packet. A bit $1$ indicates the starting point of a packet. The subsequent blocks receive the samples and the control bit-stream in parallel, and when a block sees a control bit $1$, its signal processing begins.
	
	Thanks to the accurate synchronization of the UL transmission, the \autoTrig block can indicate the accurate arrival time of packets. The subsequent blocks (e.g., CP remover, FFT) can process the UL packet at just the right time. We evaluated the performance of the \autoTrig mechanism in ROFA in Section \ref{subsec:exp_autotri_perf}.
	
	\subsection{UL CFO Estimation and Compensation}  \label{subsec:ul_cfo_est_comp}
	In conventional 802.11 infrastructure networks, the STF in the preamble is not only used for packet detection but also used for coarse CFO estimation and compensation. Since ROFA UL does away with the STF, a new CFO estimation and compensation mechanism is needed.
	
	In ROFA UL, multiple users transmit their UL packets simultaneously. As mentioned in Challenge (3), different nodes (including the AP and users) use different clocks/oscillators. Hence, there are two types of CFOs: (1) the CFO between a user and the AP; (2) the relative CFO between each pair of users. Let ${f_{{\rm{AP}}}}$ denote the carrier frequency of the AP and ${f_i}$ denote the carrier frequency of user $i$. The CFO between the AP and user $i$ is
	\begin{equation} 
		\Delta {f_{{\rm{AP}},\;i}} = {f_{{\rm{AP}}}} - {f_i}.
	\end{equation}
	The relative CFO between user $i$ and user $j$ is
	\begin{equation}
		\Delta {f_{i,j}} = {f_i} - {f_j}.
	\end{equation}
	
	The relative CFO between users causes their signals to be non-orthogonal, resulting in inter-carrier interference (ICI) if left uncompensated in the OFDMA system \cite{aziz_intercarrier_2010}. 
	
	In conventional OFDM systems, since the users do not transmit at the same time, ICI between signals of users is not an issue. However, there is still the CFO between the AP and the user. UL CFO can be estimated and compensated for with a common solution: the receiver first estimates the coarse CFO by computing autocorrelation on the STF and compensates for it. After that, the receiver uses pilots in each OFDM symbol of the coarsely compensated signals to track and compensate for residual CFOs \cite{terry_ofdm_2002}.
	
	However, the above OFDM solution is not viable for ROFA OFDMA UL, because (i) the STF has been removed; (ii) even more importantly, we have $N-1$ CFOs at the UL receiver with respect to the $N-1$ users and these CFOs are not the same. Even if the UL receiver could estimate all the CFOs perfectly without STF, the effect of CFO cannot be compensated for completely---a perfect CFO compensation for one user will induce larger CFOs for the other users.
	
	To solve this problem, ROFA UL adopts a transmitter-side solution: CFO precoding at the transmitter side, to compensate for the UL CFO before the UL transmission. That is, if a user precodes an UL packet with the inverse of the UL CFO between the AP and itself (i.e., $ - \Delta {f_{{\rm{AP}},\;i}}$), by the time the UL packet arrives at the UL receiver, the UL CFO of this packet $\Delta {f_{{\rm{AP}},\;i}}$ has already been compensated.
	
	In order to obtain $\Delta {f_{{\rm{AP}},\;i}}$, we exploit the ``reciprocity'' between DL and UL CFOs. For a radio node that uses one RF oscillator for both up-conversion and down-conversion, the up-converted transmit signals and the down-converted received signals experience the same magnitude but opposite sign CFOs. We name this reciprocity property \textbf{\textit{CFO reciprocity}}. CFO reciprocity can be represented by the following formula:
	\begin{equation}
		\Delta {f_{{\rm{AP}},\;i}} =  - \Delta {f_{i,{\rm{AP}}}},
	\end{equation}
	where $\Delta {f_{i,{\rm{AP}}}} = {f_i} - {f_{{\rm{AP}}}}$.
	
	Our experiment shows that the CFO remains approximately constant over several rounds of DL-UL phases, and that the CFOs of the DL and UL are equal in magnitude with opposite signs. We define time over which the CFO remains constant as the \textbf{\textit{CFO coherence time}}. As long as the CFO coherence time is larger than the duration of a time frame, and if the DL CFO estimation is accurate, the UL precoding can compensate for the UL CFO by virtue of reciprocity.
	
	\subsubsection{DL CFO Estimation} \label{subsubsec:dl_cfo_est}
	Recall that ROFA DL packets have a similar structure as conventional OFDM packets, in which both STF and LTF are embedded in the preamble. User $i$ can estimate the coarse CFO and the fine CFO in ROFA DL by computing autocorrelation on the STF and cross-correlation on the LTF. To estimate DL CFO with high accuracy, we place a P-LTF at the end of a DL packet (as mentioned in Section \ref{subsec:dl_phase}) and put forth an \textit{\textbf{S}TF+\textbf{L}TF+\textbf{P}-LTF (SLP) CFO estimation} method that uses P-LTF. The estimation process is elaborated below.
	
	\textbf{\textit{Coarse CFO estimation.}} In a non-highly mobile industrial setting (i.e., machines do not move at very high speeds), CFOs due to Doppler effects are not a concern. In this setting, CFOs are mainly caused by clock asynchronies between the two independent oscillators of the AP and the user.
	 
	 Let $x[n]$ denote the complex samples transmitted by the AP and $y_i[n]$ denote the complex samples received by user $i$. The CFO effect on the received signal can be written\footnote{In a practical system, CFO is not the only effect that distort the signal, other effects such as noise and channel fading also distort the signal. In this subsection, we only focus on the CFO effect.} as
	 \begin{equation} \label{eqa:cfo_effect}
	 	{y_i}[n] = x[n]{e^{j2\pi \Delta {f_i}n}}.
	 \end{equation}
 	
 	Since the STF in the DL packet is a composition of $10$ Short Training Sequences (STS) (each has $16$ samples), the DL receiver can perform autocorrelation on the periodic STSs for coarse CFO estimation. Suppose that the number of samples of one STS is $\delta $. The received $p$th STS can be written as
 	\begin{equation}	\label{eqa:p_sts_cfo}
 		{y_i}[n + p\delta ] = x[n + p\delta ]{e^{j2\pi \Delta {f_i}(n + p\delta )}},
 	\end{equation}
 	where $0 \le p \le 9$ and $x[n + p\delta ]$ is the $p$th transmitted STS. Because all the STSs in the STF are identical at the transmitter, we have
 	\begin{equation}	\label{eqa:all_sts}
 		x[n + p\delta ] = x[n].
 	\end{equation}
 	From \eqref{eqa:cfo_effect}, \eqref{eqa:p_sts_cfo}, and \eqref{eqa:all_sts}, the autocorrelation of the $p$th received STS of user $i$ can be written as
 	\begin{align} \label{eqa:p_sts_auto}
 		{z_{i,p'}} &= \sum\limits_{n' = 0}^{\delta  - 1} {y_i^*\left[ {n' + \left( {p' - 1} \right)\delta } \right]{y_i}\left[ {n' + p'\delta } \right]} \notag \\
 		&\mathop  = \sum\limits_{n' = 0}^{\delta  - 1} {{x^*}[n']{e^{ - j2\pi \Delta {f_i}\left( {n' + \left( {p' - 1} \right)\delta } \right)}}x[n']{e^{j2\pi \Delta {f_i}\left( {n' + p'\delta } \right)}}} \notag \\
 		&\mathop  = \sum\limits_{n' = 0}^{\delta  - 1} {{{\left| {x[n']} \right|}^2}{e^{j2\pi \Delta {f_i}\delta }}}
 	\end{align}
 	where $1 \le p' \le 9$. Hence, the coarse estimated CFO value of user $i$ estimated by the $p'$th STS is: 
 	\begin{align} \label{eqa:cfo_comp}
 		{\Delta \tilde f_{i,p'}} &= \frac{{\angle ({z_{i,p'}})}}{{2\pi \delta }} \notag \\
 		&\mathop  = \frac{{\angle \left( {\sum\limits_{n' = 0}^{\delta  - 1} {y_i^*\left[ {n' + \left( {p' - 1} \right)\delta } \right]{y_i}\left[ {n' + p'\delta } \right]} } \right)}}{{2\pi \delta }},
 	\end{align}
 	where $\angle ( \cdot )$ is the angle of the argument. We remark that $\Delta {\tilde f_{i,p'}}$ is the coarse estimated CFO \textit{per sample}.
 	
 	To smooth out the fluctuation of the CFO estimated in (\ref{eqa:cfo_comp}), we take an average over the coarse estimated CFO and output an average coarse estimated CFO: 
 	\begin{equation}	\label{eqa:average_coarse_cfo}
 		\overline {\Delta {{\tilde f}_{i}}}  = \frac{1}{9} \cdot \sum\limits_{p'  = 1}^9 {\frac{{\angle \left( {\sum\limits_{n'  = 0}^{\delta  - 1} {y_i^*[n'  + \left( {p'  - 1} \right)\delta ]{y_i}[n'  + p' \delta ]} } \right)}}{{2\pi \delta }}}.
 	\end{equation}
 
 	CFO estimation using the STF is not sufficiently accurate for UL CFO precoding. To improve CFO estimation accuracy, we use the LTF and P-LTF in the packet for fine CFO estimation, as elaborated below.
 	
 	\textit{\textbf{Fine CFO estimation.}} Before fine CFO estimation that uses LTF and P-LTF, user $i$ first compensates the received signal by the estimated coarse CFO above:
 	\begin{equation}
 		{r_i}[n] = {y_i}[n] \cdot {e^{ - j2\pi \overline {\Delta {{\tilde f}_i}} n}},
 	\end{equation}
 	where ${r_i}[n]$ is the coarse-compensated received signal. The fine CFO estimation may incorrectly estimate the CFO if the residual CFO rotates more than $2 \pi$ in one LTS, and that is the reason we need to first compensate the signal by the coarse CFO. The fine CFO is estimated by correlating two LTSs in the LTF:
 	\begin{equation}	\label{eqa:fine_cfo_est}
 		\Delta {\tilde f_{i,{\rm{F}}}} =  - \frac{{\angle \left( {\sum\limits_{n = 0}^{\gamma  - 1} {\underbrace {r_i^*[n]}_{{\rm{LTS1}}}\underbrace {{r_i}[n + \lambda ]}_{{\rm{LTS2}}}} } \right)}}{{2\pi \lambda }},
 	\end{equation}
 	where $\gamma $ is the number of samples in one LTS, $\lambda$ is the distance in terms of the number of samples between the first samples of the two LTSs. In \eqref{eqa:fine_cfo_est}, $n=0$ corresponds to the first sample of the first LTS. ${r_i}[n]$ and ${r_i}[n + \lambda ]$ are the first received LTS (LTS1) and the second received LTS (LTS2) respectively. Since LTS1 and LTS2 are adjacent to each other in the LTF, the distance between them is equal to the length of one LTS, i.e., $\lambda  = \gamma $.
 	
 	The P-LTF embedded at the end of the DL packet can be used to further improve the accuracy of fine CFO estimation. We name the step of CFO estimation that uses P-LTF \textit{Ultra-Fine CFO estimation}.
 	
 	\textit{\textbf{Ultra-fine CFO estimation.}} Before ultra-fine CFO estimation, the receiver compensates for the fine CFO estimated in (\ref{eqa:fine_cfo_est}):
 	\begin{equation} \label{eqa:fine_cfo_comp}
 		r'_i[n] = {r_i}[n] \cdot {e^{ - j2\pi \Delta {{\tilde f}_{i,{\rm{F}}}}n}}.
 	\end{equation}
 	Since the distance between P-LTF and LTS1 is ${\lambda _{\rm{P}}} = {N_{{\rm{data}}}} + \gamma $. The ultra-fine CFO estimation is computed by
 	\begin{equation} \label{eqa:ultrafine_cfo_est}
 		\Delta {\tilde f_{i,{\rm{UF}}}} =  - \frac{{\angle \left( {\sum\limits_{n = 0}^{\gamma  - 1} {\underbrace {{{\left( {r{'_i}[n]} \right)}^*}}_{{\rm{LTS1}}}\underbrace {r{'_i}[n + {\lambda _{\rm{P}}}]}_{{\rm{P - LTF}}}} } \right)}}{{2\pi {\lambda _{\rm{P}}}}}.
 	\end{equation}
 
 	In \eqref{eqa:ultrafine_cfo_est}, the P-LTF CFO estimation reduces the noise in the correlation by increasing the distance between the two components in correlation. Specifically, the distance has been increased by $\left( {{N_{{\rm{data}}}} + \gamma } \right)/\gamma $ times than that of \eqref{eqa:fine_cfo_est}, resulting in a more accurate CFO estimation.
 	
 	\begin{figure}
 		\centering {\includegraphics[width=\oneFigWidth]{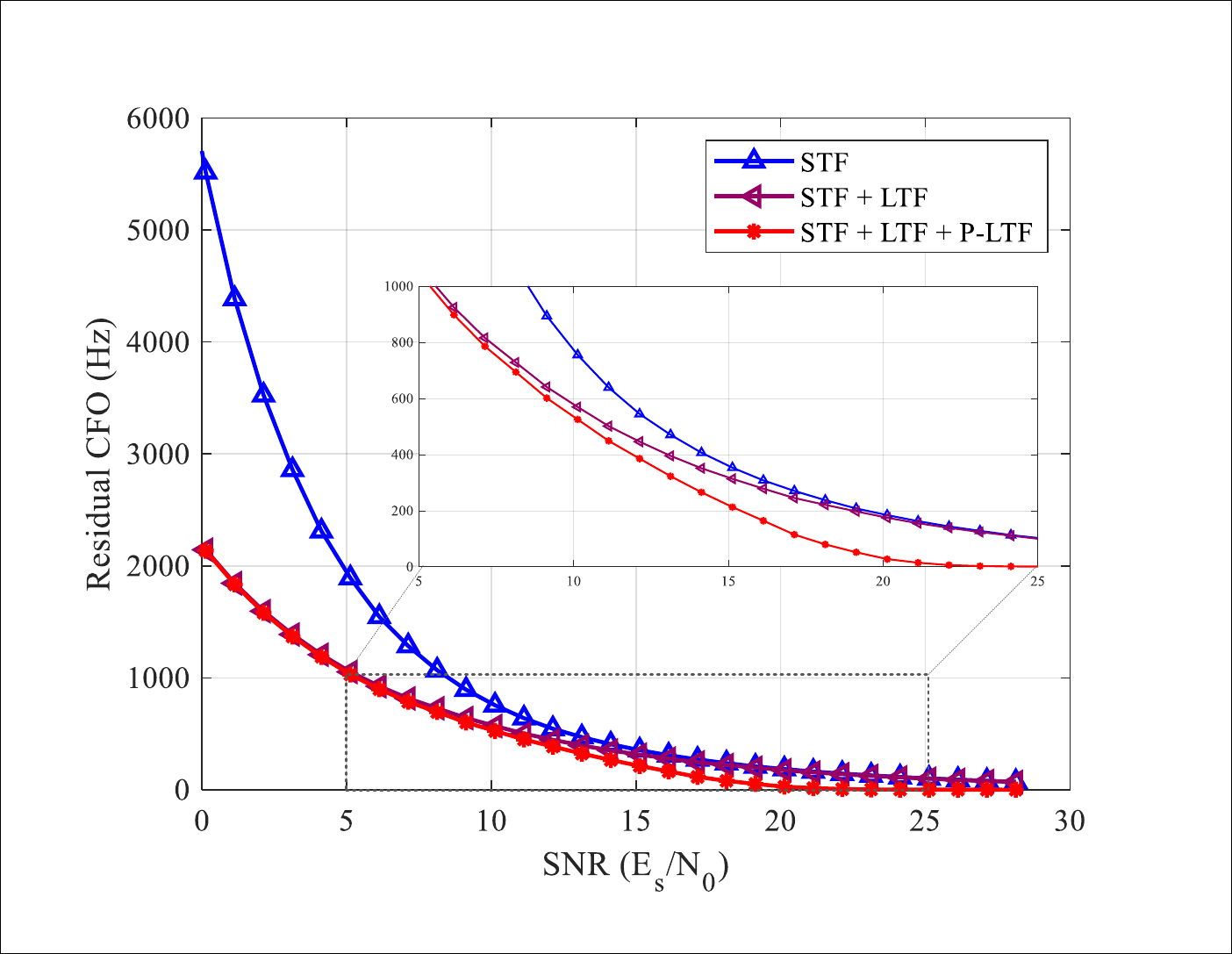}}
 		\caption{The residual CFO after DL CFO estimation and compensation of different methods: (1) using STF only, (2) using STF and LTF, (3) using STF, LTF, and P-LTF.}
 		\label{fig:pltf_residual}
 	\end{figure}
 
 	We ran a quick simulation to investigate residual DL CFO versus Signal-to-Noise Ratio (SNR) after compensating for the DL CFO that uses different components (STF, LTF, or P-LTF) of the packet. We ran ${10^5}$ simulations, one for one packet. The residual DL CFO for packet $k$ is computed by
 	\begin{equation}	\label{eqa:residual_cfo_comp}
 		\Delta {\tilde \theta _{k,{\rm{res}}}} = \left| {\Delta {{\tilde \theta }_k} - \Delta f} \right| ,
 	\end{equation}
 	where $\Delta {\tilde \theta _k}$ is the final estimated CFO of packet $k$ (using a combination of STF, LTF, or P-LTF, depends on different methods), and $\Delta f$ is the ground truth DL CFO. Because the estimation method is the same for all users, we omit the user index $i$ in (\ref{eqa:residual_cfo_comp}).
 	
 	 For our ${10^5}$ simulations, we fixed $\Delta f$ to be $0.004\pi \;{\rm{rad/sample}}$. We found the results to be similar for different $\Delta f$. Note that the unit of the variables in (\ref{eqa:residual_cfo_comp}) are all in $\rm{rad/sample}$. The conversion between $Hz$ and $\rm{rad/sample}$ for a $10MHz$ system is defined as follows:
 	 \begin{equation} \label{eqa:hertz_conversion}
 	 	1\;{\rm{rad/sample}} = {10^7}\;{\rm{rads/}}\sec  = \frac{{{{10}^7}}}{{2\pi }}Hz.
 	 \end{equation}
  	Thus, $\Delta f$ in the unit of Hertz is $\Delta {f^{{\rm{(Hz)}}}} = \Delta f \cdot ({10^7}/2\pi ) = 20KHz$. The final results $\Delta \tilde \theta _{k,{\rm{res}}}^{{\rm{(Hz)}}}$, where $\Delta \tilde \theta _{k,{\rm{res}}}^{{\rm{(Hz)}}} = \Delta {\tilde \theta _{k,{\rm{res}}}}({10^7}/2\pi )$, have been averaged over $10^5$ tests and are shown in Fig. \ref{fig:pltf_residual}.
  	
  	From Fig. \ref{fig:pltf_residual} we can see that STF+LTF+P-LTF (SLP) CFO estimation always outperforms STF CFO estimation. On the other hand, the improvement of SLP CFO estimation over STF+LTF CFO estimation is noticeable in the middle to high SNR regime but negligible in the low SNR regime. In the following, we explain the lack of improvement in the low SNR regime and provide a remedy to ensure our proposed method also works well in the low SNR regime.
  	
\begin{figure}
  		\centering {\includegraphics[width=\oneFigWidth]{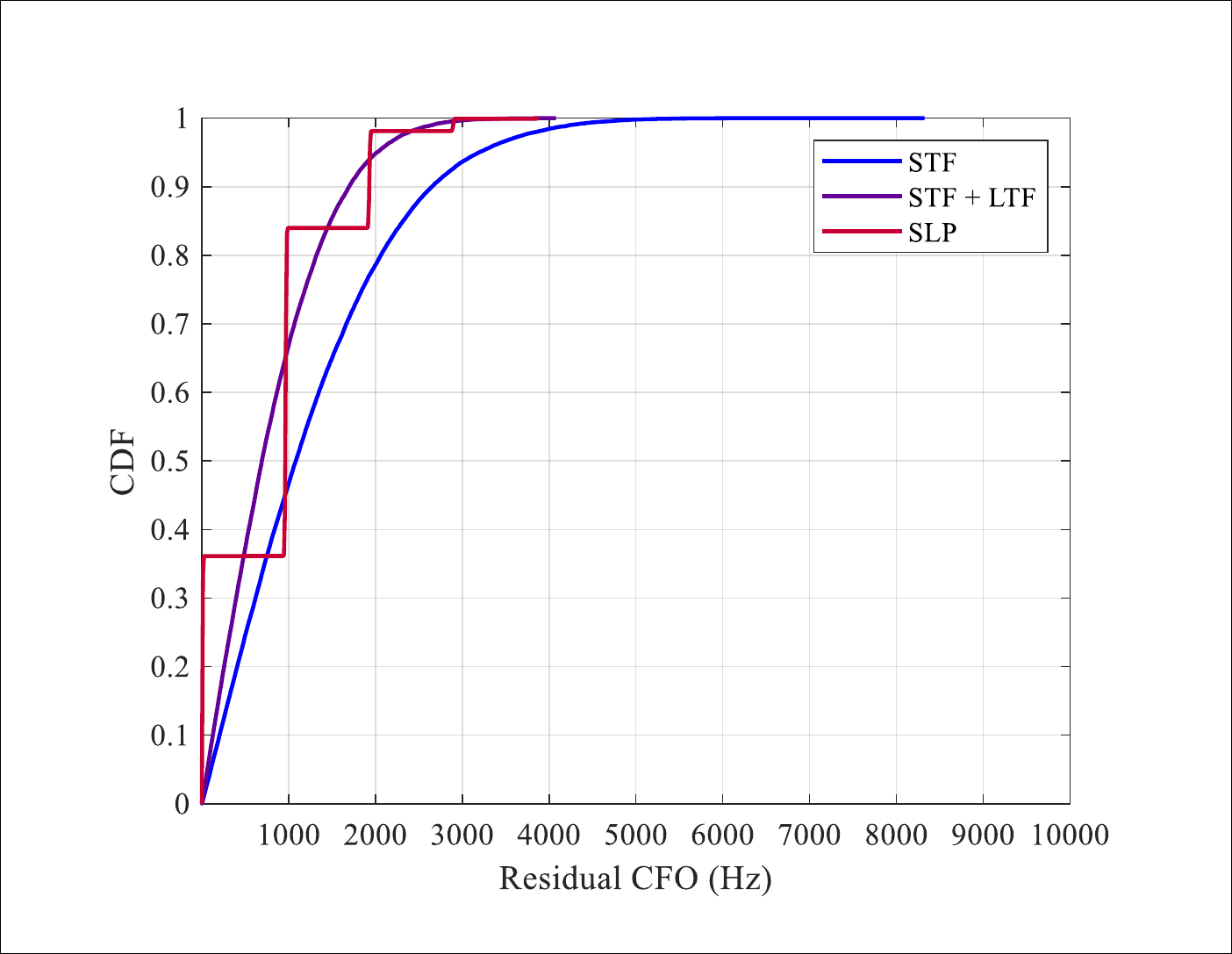}}
  		\caption{The CDF of $\Delta \tilde \theta _{k,{\rm{res}}}^{{\rm{(Hz)}}}$ of three DL CFO estimation methods with packet length is equal to $128$ OFDM symbols.}
  		\label{fig:ecdf_cfo_est}
  	\end{figure}

  	\textit{\textbf{Ultra-fine CFO Estimation with Middle-LTF.}} To investigate the cause of the estimation degradation of STF+LTF+P-LTF CFO estimation in low SNR regime, we get the cumulative distribution function (CDF) of $\Delta \tilde \theta _{k,{\rm{res}}}^{{\rm{(Hz)}}}$ of three different methods from the same simulation with SNR set to $10dB$, and plot it in Fig. \ref{fig:ecdf_cfo_est}.
  	
  	From Fig. \ref{fig:ecdf_cfo_est} we can see that the CDF of $\Delta \tilde \theta _{k,{\rm{res}}}^{{\rm{(Hz)}}}$ of SLP CFO estimation has a ``staircase'' shape in the low SNR regime. We believe that when the SNR is low, the CFO estimation that uses STF and LTF is unreliable, causing the phase of the signal to rotate more than $2\pi$ from the beginning of the packet to the end of the packet, even after compensation using the estimated CFO. As a result, the ultra-fine CFO estimation using LTF and P-LTF, which assumes the phase rotation is less than $2\pi$, runs into problems. We provide the detailed analysis in Appendix \ref{append_analy_residual_cfo}.
  	
\begin{figure}
   	\centering {\includegraphics[width=\structFigWidth]{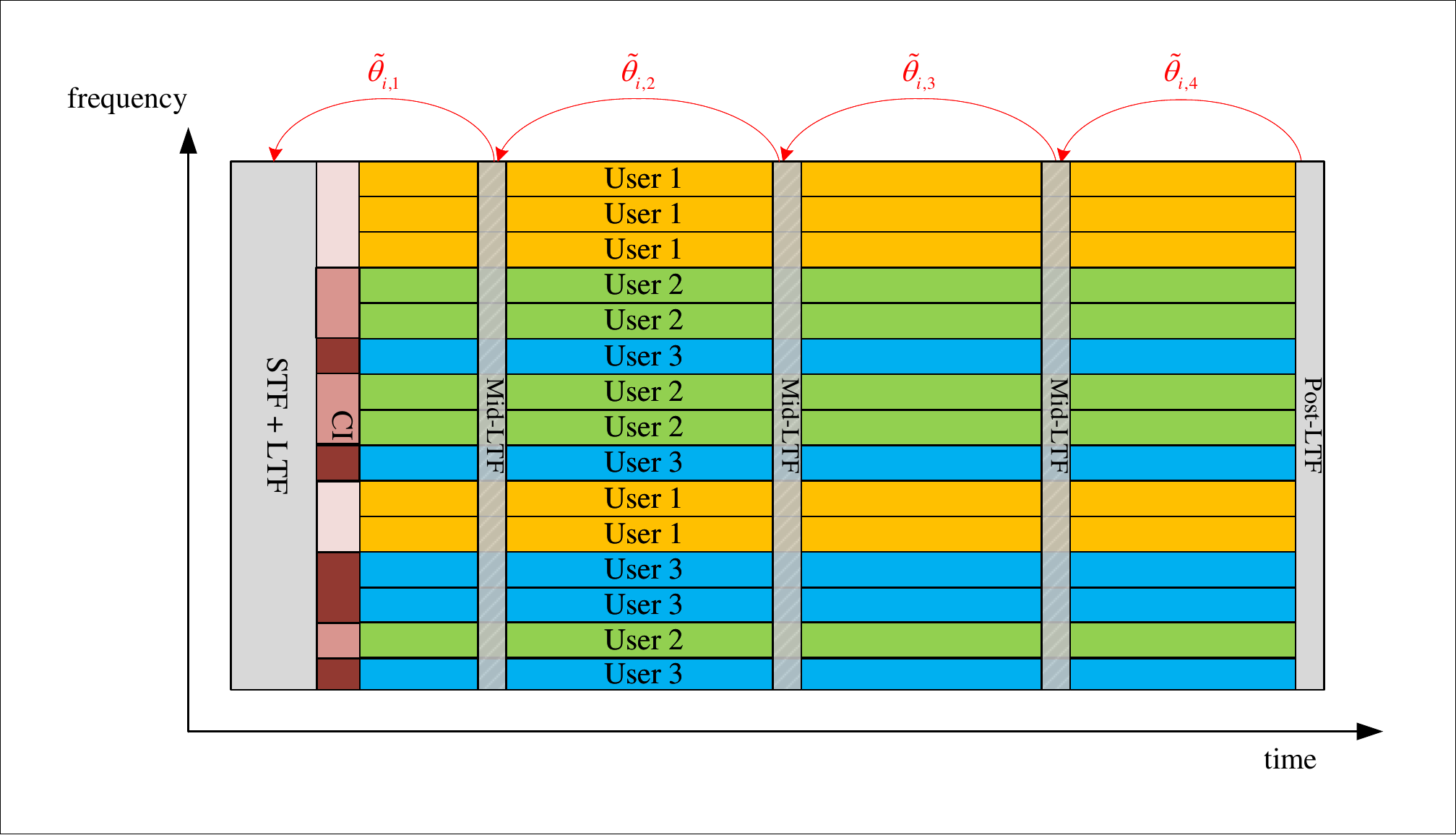}}
   	\caption{The structure of a DL packet with three mid-LTFs.}
   	\label{fig:mid_ltf}
   \end{figure}

  	To solve the above problem in the low-SNR regime, we propose to reduce the gap of computing the correlation between two LTFs. Specifically, we insert ${N_{{\rm{M - LTF}}}}$ Middle-LTFs (M-LTF) into the packet with a constant gap ${\lambda _{\rm{M}}}$. An example of inserting $3$ M-LTFs is shown in Fig. \ref{fig:mid_ltf}. One M-LTF contains one LTS.

  	Suppose the actual phase rotation from the LTF to the P-LTF is
  	\begin{equation}	\label{eqa:cfo_combine}
  		{\theta _{{\rm{L}} \to {\rm{P}}}} = {\theta '_{{\rm{L}} \to {\rm{P}}}} + m \cdot 2\pi,
  	\end{equation}
  	 where ${\theta '_{{\rm{L}} \to {\rm{P}}}}$ is the factional part of the residual CFO and $m \cdot 2\pi $ is the integer part. When $m>0$, only the factional part of the residual CFO can be measured by correlating LTF and P-LTF. In order to recover the integer part of the CFO, M-LTFs are used in ultra-fine CFO estimation to find out $m$ in \eqref{eqa:cfo_combine}. The recovering procedures are as follow:
  	 
  	 After the fine CFO compensation that uses STF+LTF (i.e., the process in \eqref{eqa:fine_cfo_comp}), the DL receivers compute the phase rotation between each pair of adjacent M-LTFs (see the red curves in Fig. \ref{fig:mid_ltf}). The estimated phase rotation between the $j$th M-LTF and the $(j-1)$th M-LTF can be written by
  	 \begin{equation}	\label{eqa:midamble_correlate}
  	 	{\tilde \theta _{i,j}} =  - \angle \left( {\sum\limits_{n = 0}^{\gamma  - 1} {\underbrace {{{\left( {r'_i [n + {\lambda _{j - 1,{\rm{M}}}}]} \right)}^*}}_{(j - 1){\rm{th}}\;{\rm{M - LTF}}}\underbrace {r'_i[n + {\lambda _{j,{\rm{M}}}}]}_{j{\rm{th}}\;{\rm{M - LTF}}}} } \right),
  	 \end{equation}
   	where ${\lambda _{j,{\rm{M}}}}$ is the distance between LTS2 and the $j$th M-LTF. The computation of ${\tilde \theta _{i,j}}$ is done for $1 \le j \le {N_{{\rm{M - LTF}}}} + 1$. Thus, LTS2 and P-LTF are also included in computing ${\tilde \theta _{i,j}}$ when $j=1$ and $j = {N_{{\rm{M - LTF}}}} + 1$, respectively. 
   	
   	We then compute $m$ by calculating
   	\begin{equation}	\label{eqa:fraction_cfo}
   		m = \left\lfloor {\frac{{\sum\nolimits_1^{{N_{{\rm{M - LTF}}}} + 1} {{{\tilde \theta }_{i,j}}} }}{{2\pi }}} \right\rfloor ,
   	\end{equation}
   	where $\left\lfloor  \cdot  \right\rfloor $ represents the floor function that outputs the greatest integer less than or equal to the given number.
   	
   	Meanwhile,  the process in (\ref{eqa:ultrafine_cfo_est}) is also carried out\footnote{We do not need to compensate the signal with the integer part of the residual CFO, because it is always $m \cdot 2\pi $.} to get
   	\begin{equation}	\label{eqa:integer_cfo}
   		{\tilde \theta '_{{\rm{L}} \to {\rm{P}}}} = \Delta {\tilde f_{i,{\rm{UF}}}} \cdot {\lambda _{\rm{P}}}
   	\end{equation}
   	Putting (\ref{eqa:fraction_cfo}) and (\ref{eqa:integer_cfo}) into (\ref{eqa:cfo_combine}), we have the final output of \textbf{U}ltra-\textbf{F}ine CFO estimation with \textbf{M}-LTF, which can be written as
   	\begin{equation}
   		\Delta {\tilde f_{i,{\rm{UFM}}}} = \frac{{{{\tilde \theta }_{{\rm{L}} \to {\rm{P}}}}}}{{{\lambda _{\rm{P}}}}} = \Delta {\tilde f_{i,{\rm{UF}}}} + \left\lfloor {\frac{{\sum\nolimits_1^{{N_{{\rm{M - LTF}}}} + 1} {{{\tilde \theta }_{i,j}}} }}{{2\pi }}} \right\rfloor \frac{{2\pi }}{{{\lambda _{\rm{P}}}}}.
   	\end{equation}
   
   \textbf{\textit{Remark}}: The determination of ${\lambda _{\rm{M}}}$ is based on the frequency stability of the oscillator in the RF hardware. The smaller ${\lambda _{\rm{M}}}$, the larger the frequency instability that ultra-fine CFO estimation can handle. However, this capability comes at the cost of more overhead. In our experiments, ${\lambda _{\rm{M}}}$ equal to $32$ OFDM symbols is sufficient for a system operating at $10MHz$.
     	
   \begin{figure}
   	\centering {\includegraphics[width=\figWidth]{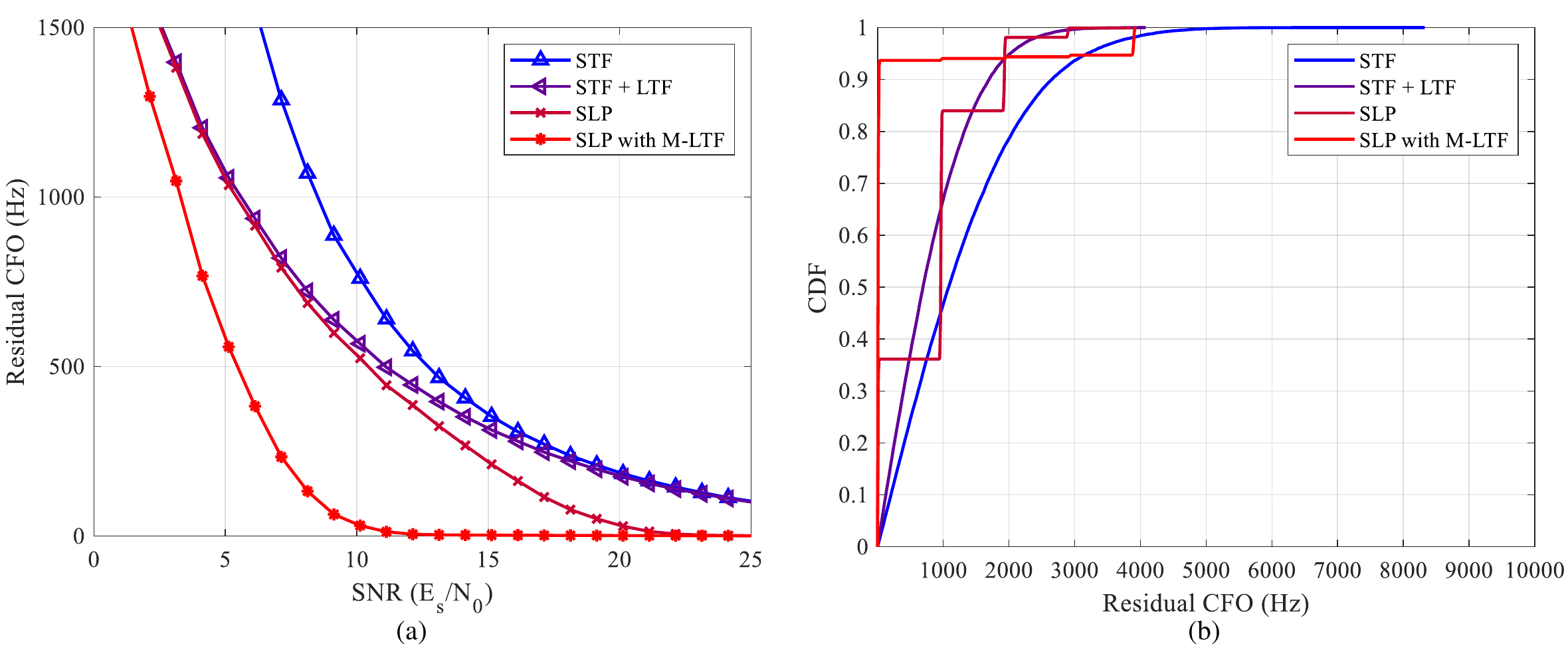}}
   	\caption{The performance of ultra-fine CFO estimation when involving the M-LTFs: (a) Average residual CFO versus SNR; (b) CDF of residual CFO at $SNR=10dB$.}
   	\label{fig:mid_stf_perf}
   \end{figure}

	The performance of using M-LTF in the ultra-fine CFO estimation is shown in Fig. \ref{fig:mid_stf_perf}. The residual CFO after DL CFO estimation is significantly improved by SLP CFO estimation with M-LTFs. The DL receivers can achieve nearly perfect CFO estimation at around $12dB$, compared to the one without M-LTFs at $21dB$. At a typical SNR regime, e.g., $10dB$, the percentage of perfect CFO estimation (residual CFO $ \approx 0$) increases from $35\%$ to $92\%$.
	
	\subsubsection{UL CFO Precoding} \label{subsubsec:ul_cfo_prec}
  	After estimating the CFO in the DL transmission, user $i$ precodes its UL packet with a phase ${\theta _i}$ for UL transmissions. Specifically, ${\theta _i}$ is obtained from the SLP CFO estimation method:
  	\begin{equation}
  		{\theta _i} =  - (\underbrace {\overline {\Delta {{\tilde f}_i}} }_{{\rm{Coarse}}} + \underbrace {\Delta {f_{i,{\rm{F}}}}}_{{\rm{Fine}}} + \underbrace {\Delta {f_{i,{\rm{UFM}}}}}_{{\rm{Ultra - Fine}}}).
  	\end{equation}
  	Suppose the UL signal of user $i$ is ${x_{i,{\rm{UL}}}}[n]$, the precoded signal ${x'_{i,{\rm{UL}}}}[n]$ can be written as
  	\begin{equation}
  		{x'_{i,{\rm{UL}}}}[n] = {x_{i,{\rm{UL}}}}[n] \cdot {e^{ - j2\pi \left( {\overline {\Delta {{\tilde f}_i}}  + \Delta {f_{i,{\rm{F}}}} + \Delta {f_{i,{\rm{UFM}}}}} \right)n}}.
  	\end{equation}
  	For the preparation of an UL packet, the precoder precodes the UL packet after the OFDMA modulation on its time-domain signal. The system-level performance of the UL CFO compensation in terms of error vector magnitude (EVM) \cite{mckinley_2004_evm}, bit-error rate (BER) and packet-error rate (PER) will be detailed in Sections \ref{subsec:exp_ul_prec_perf} and \ref{subsec:exp_reliability}.
	
	\subsection{UL Channel Estimation and Equalization} \label{subsec:UL_channel_est_equal}
	Although STF is omitted, LTF is still kept in the UL packet for channel estimation and equalization at the receiver. Let ${y_{\rm{UL}}}(t)$ be the signal received by the UL receiver (AP). The received UL signals, which are superimposed signals of many users, can be written as
	\begin{equation}
		{y_{\rm{UL}}}(t) = \sum_{i = 1}^{N - 1} {({h_{i,\rm{UL}}} * {x'_{i,\rm{UL}}})(t) + n_{i,\rm{UL}}(t)},
	\end{equation}
	where ${h_{i,\rm{UL}}}(t)$ is the channel coefficient of the UL channel between user $i$ and the AP, $n_{i,\rm{UL}}(t)$ is the white Gaussian noise with zero mean and variance ${\sigma ^2}$, and $*$ is the convolution operation.
	
	For channel equalization in ROFA UL, the AP first estimates the UL channel of each user separately. User $i$ generates an LTF and appends it to the beginning of the UL packet. Specifically, the LTF is generated in the frequency domain. In user $i$'s LTF ${L_i}$, a complex symbol on the $m$-th subcarrier is generated by
	\begin{equation} \label{eqa:ul_ltf}
		{L_i}[m] = \left\{ {\begin{array}{*{20}{c}}
				{l,}&{m \in {\cal S}_i^{}}\\
				{0,}&{m \notin {\cal S}_i^{}}
		\end{array}} \right. ,
	\end{equation}
	where $l$ is a BPSK symbol sampled from a random BPSK generator. Only the subcarriers allocated to user $i$ will be inserted with a BPSK symbol, other subcarriers are filled with $0$s. The LTF ${L_i}$ then forms the input to the IFFT block. The output of IFFT, a time-domain LTF, is appended to the beginning of user $i$'s UL packet. Since the subcarrier allocations for different users are orthogonal, the generation function (\ref{eqa:ul_ltf}) guarantees that the LTFs generated by different users are not overlapped in the frequency domain.
	
	At the AP's receiver side, the AP already has knowledge of the subcarrier allocation ${\cal S}_i$ of user $i$. When the \autoTrig block indicates the arrival of an UL packet, the reception process kick starts by the  AP forwarding the signal to the FFT block (see Fig. \ref{fig:flowgraph}). From the output of the FFT, the AP extracts the channel coefficients for user $i$ by
	\begin{equation}
		{\tilde H_i}[m] = \frac{1}{2} \cdot \left( {\frac{{{{\hat L}_{i,1}}[m]}}{{{L_i}[m]}} + \frac{{{{\hat L}_{i,2}}[m]}}{{{L_i}[m]}}} \right),\;\;\;m \in {\cal S}_i,
	\end{equation}
	where ${\hat L_{i,1}}[m]$ is the first received LTS, ${\hat L_{i,2}}[m]$ is the second received LTS. With the above ${\tilde H_i}[m]$, the AP can equalize the channel and provide the equalized complex symbols for the QAM demodulation block.
	
	\section{Reliability of ROFA: Effective SNR Improvement} \label{sec:reliability_rofa}
	From the hardware perspective, the largest transmit power of a wireless device is constrained by its RF components, including power amplifier and Analog-to-Digital Converter (ADC)\footnote{For a power amplifier, there are linear region as well as non-linear region for power amplifying. In our paper, we only consider the power amplifying in the linear region.}. The largest transmit power of a wireless device at any moment cannot exceed a certain threshold if it wants to operate in the linear-amplification regime.
	
	Each user in ROFA UL only picks a subset of subcarriers to transmit signals. Since the maximum transmit power of a wireless device is fixed at ${P_{\max }}$, the fewer subcarriers a transmitter use, the larger the power a transmitter can concentrate on each of the used subcarriers. For a $52$-subcarrier ($48$ data subcarriers and $4$ pilots) $64$-FFT OFDM system, each subcarrier on average takes up $\frac{1}{{52}}{P_{\max }}$ for its transmission. For comparison, for a user using only three subcarriers in a $64$-FFT OFDMA system, each subcarrier can have $\frac{1}{3}{P_{\max }}$ for the transmission. From the perspective of transmit power on the occupied subcarriers, the OFDMA system can have $52/3$ times larger power than the OFDM system. If they are transmitting over the same channel, the effective received SNR will be increased by $12.38dB$ if the OFDMA system is adopted.
	
	\section{Experimental Validation} \label{sec:experiment}
	\subsection{Experimental Setup} \label{subsec:exp_setup}
	In this section, we present the experimental validation of ROFA. In the experiment, five radio nodes are placed in an indoor office environment. Each radio node includes a PC and a USRP. To explore the compatibility of ROFA with hardware, we select different types of USRPs for the radio nodes. Three of them are USRP X310s and two of them are USRP N210s. One of the USRP X310 is chosen to be the AP.
	
	\begin{figure}
		\centering {\includegraphics[width=\oneFigWidth]{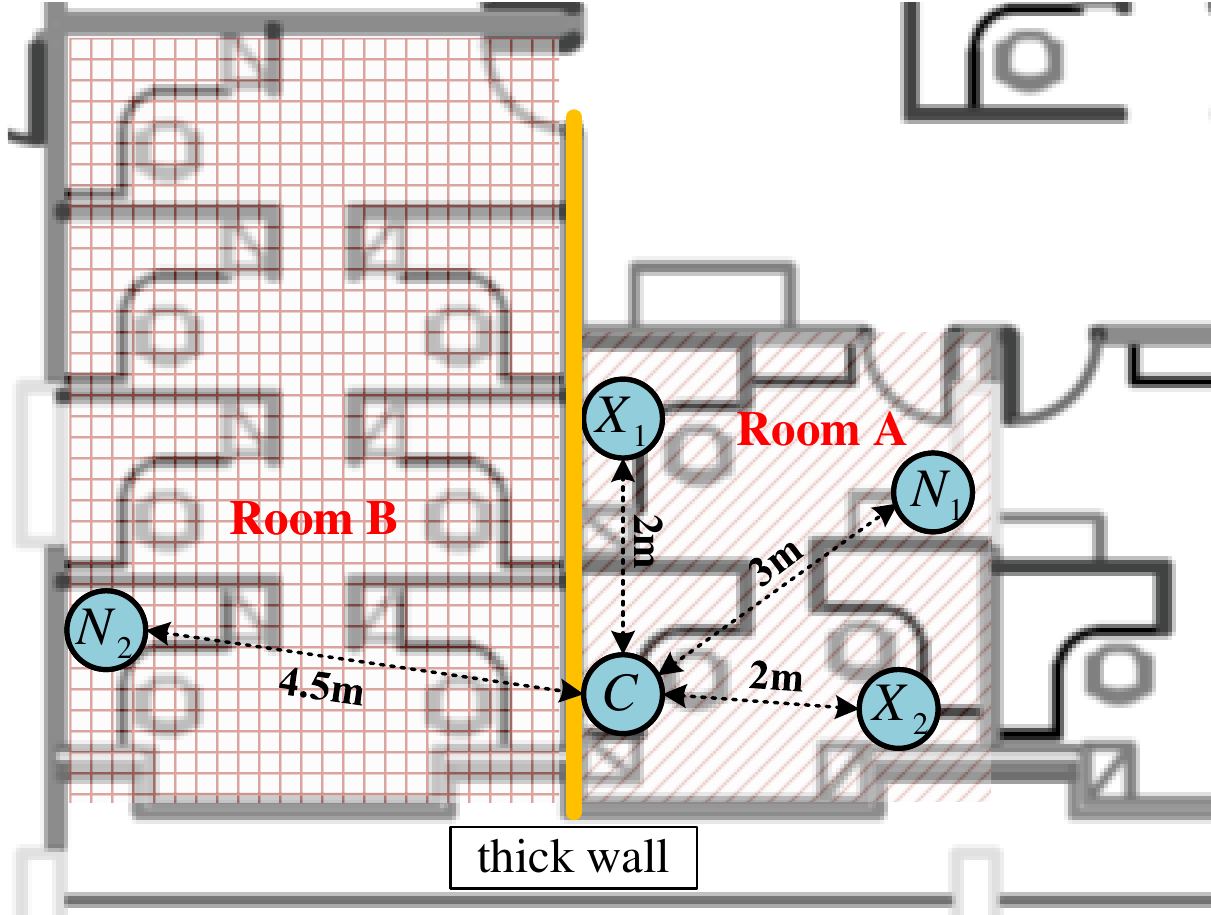}}
		\caption{The deployed locations of the radio nodes in the office. $C$ is the AP (USRP X310). ${X_1},{X_2}$ are users running on USRP X310s, and ${N_1},{N_2}$ are users running on USRP N210s.}
		\label{fig:node_deployment}
	\end{figure}
	
	The floor plan and the locations of the radio nodes are shown in Fig. \ref{fig:node_deployment}. Four radio nodes (three USRP X310s ($C,{X_1},{X_2}$) and one USRP N210 (${N_1}$)) are deployed in Room A. Another USRP N210 is deployed in Room B. Room A and Room B are next to each other but separated by a thick wall (the yellow line in Fig. \ref{fig:node_deployment}) The size of Room A is around $4$ meters by $4$ meters, while the size of Room B is $8$ meters by $5$ meters. One of the radio nodes with USRP X310, node $C$ in Room A, serves as the AP throughout the experiments. The AP is approximately equidistant to two of the nodes ${X_1},{X_2}$ in the same room ($2$ meters). Another node in the same room ${N_1}$ is $3$ meters away from the AP. In addition to four nodes in Room A, a radio node ${N_2}$ is placed in Room B, $4.5$ meters away from the AP. Because Room A and Room B are separated by a thick wall, the communication between the AP and ${N_2}$ is Non-Light-of-Sight (NLoS). Meanwhile, the communication channels $C \leftrightarrow {X_1}$, $C \leftrightarrow {X_2}$, and $C \leftrightarrow {N_1}$ are Light-of-Sight (LoS) channels.
	
	\begin{figure*}
		\centering {\includegraphics[width=\twoColumnFigWidth]{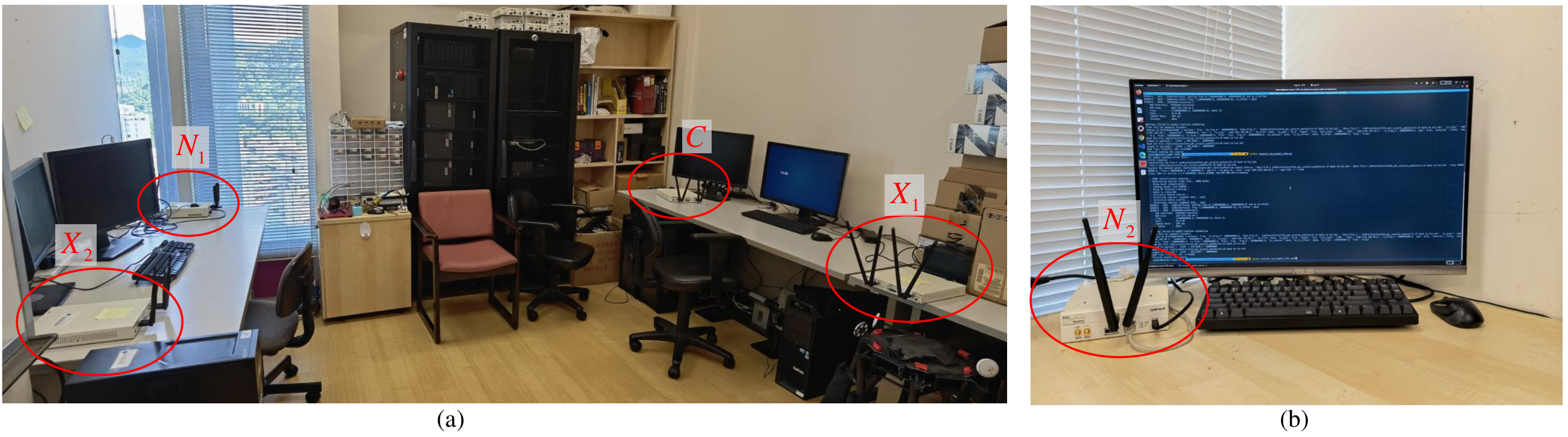}}
		\caption{The deployment of radio nodes in (a) Room A and (b) Room B.}
		\label{fig:exp_platform}
	\end{figure*}

	In the PC-based SDR platform, the USRP is responsible for transmitting and receiving signals over the air, while the PC is responsible for the preparation of the transmit baseband signal and processing of the received baseband signal. We implemented the signal processing of ROFA on GNURadio. GNURadio is an open-source software platform for real-time signal processing over a PC. All the signal processing blocks in Fig. \ref{fig:flowgraph} have been implemented on GNURadio. Specifically, the blocks are written in C++, while the flowgraphs (connections of the blocks) are written in Python.

	All the USRPs are equipped with onboard TCXOs. Each of the USRP X310s has a UBX-160 daughterboard and is connected to a PC via a 10Gbps Ethernet cable. Each of the USRP N210s has an SBX-40 daughterboard and is connected to a PC via a 1Gbps Ethernet cable. The USRP X310 of the AP is connected to a powerful PC with a 16-core AMD 1950X Processor 3.4GHz and 64G RAM. Each of the other USRPs (both X310 and N210) is connected to a PC with a 4-core i7-4790k 4.0GHz and 32G RAM. The operating system is Ubuntu 18.04 LTS with kernel 4.15. USRP Hardware Driver (UHD) 3.9.7 and GNURadio 3.7.11 are installed in the operating system for USRP control and signal processing.
	
	\begin{table}[]
		\caption{\textcolor{black}{Parameters of the PHY-layer in the experiment}}\label{tab:exp}
		\centering
		\def\arraystretch{1}
		{\color{black}\begin{tabular}{l|c}
				\hline
				Center frequency   & $2.418GHz$               \\ \hline
				Bandwidth          & $10MHz$                  \\ \hline
				\parbox{0.4\linewidth}{Number of OFDM symbols for the payload } & \parbox{0.3\linewidth}{$128$ (ROFA), \\ $8$ (OFDM-TDMA)} \\ \hline
				Modulation         & BPSK                     \\ \hline
				Channel code       & $1/2$ convolutional code \\ \hline
				Length of cyclic-prefix (CP) & $16$ samples        \\ \hline
				Guard interval	& $80$ samples			\\	\hline
				FFT size ${N_{{\rm{FFT}}}}$  & $64$			\\ \hline
				Number of subcarriers & $3$ (ROFA), $48$ (OFDM-TDMA)			\\ \hline
		\end{tabular}}
	\end{table}
	
	In this experiment section, both ROFA and OFDM-TDMA adopt the PHY-layer parameters given in TABLE \ref{tab:exp}. Specifically, ROFA and OFDM-TDMA transmit the same amount of data in a packet. Therefore, the numbers of OFDM symbols for the payload are different in two system, because the users in two systems use different numbers of subcarriers. We also set the number of OFDM symbols for payload in the DL packet and the UL packet to be the same in the experiments. The duration of a DL packet and an UL packet can be computed as follows:
	\begin{align*}
				{T_{{\rm{DL}}}} &= \left( {320 + 128 \times 80 + 80} \right)/{10^7} \text{sample/sec}  = 1.064ms\\
				{T_{{\rm{UL}}}} &= \left( {2 \times 80 + 128 \times 80} \right)/{10^7}\text{sample/sec}  = 1.04ms.
	\end{align*}

	The TDMA guard interval in TABLE \ref{tab:exp} is the time duration between the end of the packet and the beginning of the next packet. This guard interval reduces the signal interferences to the next slot, caused by propagation delay. For our $10MHz$ system, an $80$-sample guard interval can ensure the packet in the latter time slot will not be interfered with by the former packet, with the propagation delay smaller than $8\mu s$.
	
	We next validate the assumption in the design of ROFA that the DL CFO and the UL CFO are equal in magnitudes with opposite signs. After the validation, we present the EVM performance of the UL CFO compensation at the receiver with and without the UL CFO precoding at the transmitter. Finally, we compare the BER and PER performance of ROFA (an OFDMA system) versus RTTS-SDR (an OFDM-TDMA system). We also investigate possible near-far effects in ROFA UL.
	
	\subsection{Relative CFO between Two Nodes}  \label{subsec:exp_relative_cfo}
	\begin{figure}
		\centering {\includegraphics[width=\oneFigWidth]{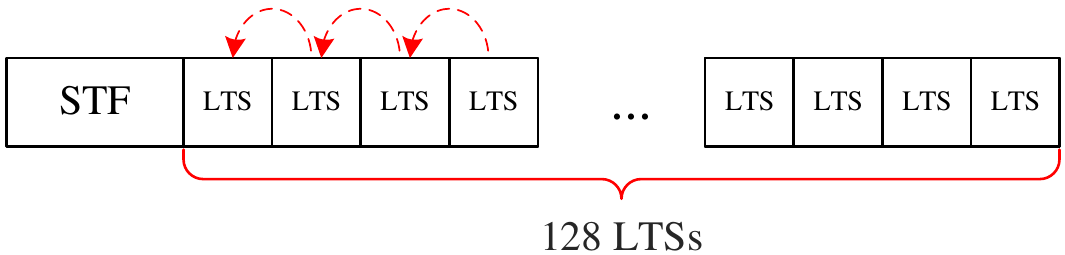}}
		\caption{The structure of a channel probe packet.}
		\label{fig:struct_channel_probe}
	\end{figure}

	A key assumption of ROFA is that the relative CFOs between two oscillators in two directions are equal in magnitudes with different signs, and they can remain unchanged for at least one round of DL-UL phases. If the assumption is valid, the user can make use of the estimated DL CFO to precode the UL packet. In this subsection, we present our measurement of the variation of the relative CFO between two nodes over time and the validation of CFO reciprocity.
	
	To measure the variation of the relative CFO between two nodes, we designed a channel probe packet, as shown in Fig. \ref{fig:struct_channel_probe}. The channel probe packet starts with an STF, followed by $128$ LTSs.

	In this experiment, one radio node (say Node A) and another radio node (Node B) take turns transmitting the channel probe packet to each other repeatedly. In other words, Node A first sent a channel probe packet to Node B. Upon receiving the channel probe packet, Node B immediately sent back a channel probe packet to Node A.
	
	For both nodes, after receiving the channel probe packet, the receiver first uses the STF to estimate and compensate for the coarse CFO. Then it continuously measures the residual relative CFO by calculating the correlation between two adjacent LTSs. To eliminate measuring errors caused by the noise (we are only interested in investigating CFO reciprocity and CFO coherence time here), the transmit power was set to a high level---the received SNR is more than $40dB$, so that the CFO estimation error was small.
	
	\begin{figure}
		\centering {\includegraphics[width=\linewidth]{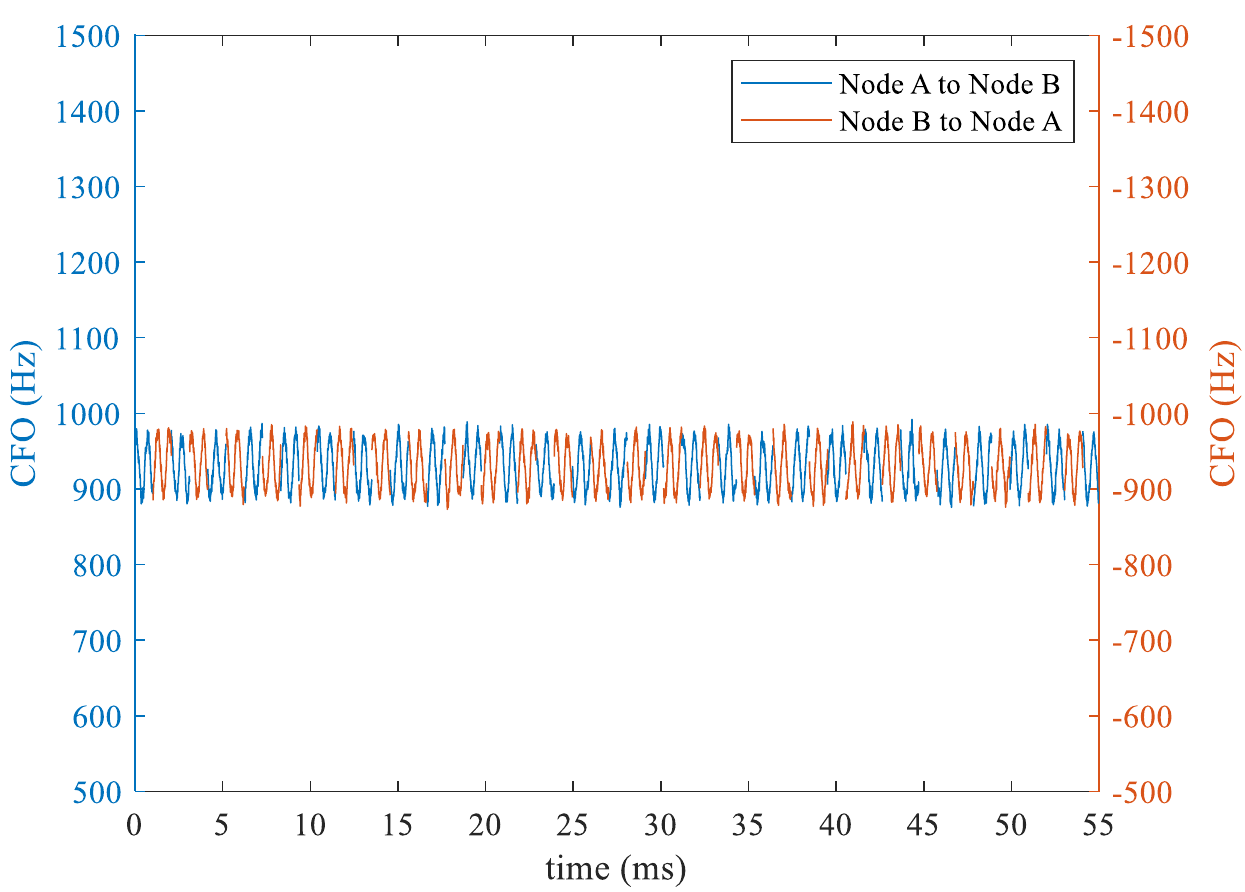}}
		\caption{Relative CFO variations of two nodes over time in two directions.}
		\label{fig:relative_cfo}
	\end{figure}

	Fig. \ref{fig:relative_cfo} shows the relative CFO variation of both directions over $55ms$, which is more than $25$ rounds of DL and UL. Taking the relative CFO measured by the probe packets from Node A to Node B (blue curve) as an example, the average relative CFO is about $930Hz$ and the variation is no larger than $110Hz$ within $55ms$ (i.e., the magnitudes of the UL and DL CFOs differ by no more than $110Hz$)  Comparing the worst variation, $110Hz$, with the subcarrier spacing ($10MHz/64 = 156.25KHz$), the UL CFO is considered to be negligible as far as the ICI between users is concerned (i.e., if after UL precoding, the maximum residual CFO of $110Hz$ is acceptable from the ICI standpoint). 
	
	\subsection{Performance of Auto-trigger} \label{subsec:exp_autotri_perf}
	This subsection presents the performance validation of replacing the packet detection block and time synchronizer block with the \autoTrig block. Let $t_{{\rm{auto}}}^k$ be the $k$th UL packet's targeted arrival time provided by \autoTrig and $t_{{\rm{actual}}}^k$ be the actual arrival time measured. We use $\left( {t_{{\rm{auto}}}^k - t_{{\rm{actual}}}^k} \right)$ as the metric of Auto-trigger's performance.
	
	To measure the actual arrival time of an UL packet, we put the conventional STF and the conventional LTF back in the UL packet and let only one user transmit an UL packet at a time. At the UL receiver side, we create two receive paths: (1) a conventional receive path that uses the conventional STF auto-correlation + LTF cross-correction for packet detection that triggers the subsequent receiver signal processing; (2) a receive path that instead uses the \autoTrig block for the triggering.
	
	\begin{figure}
		\centering {\includegraphics[width=\oneFigWidth]{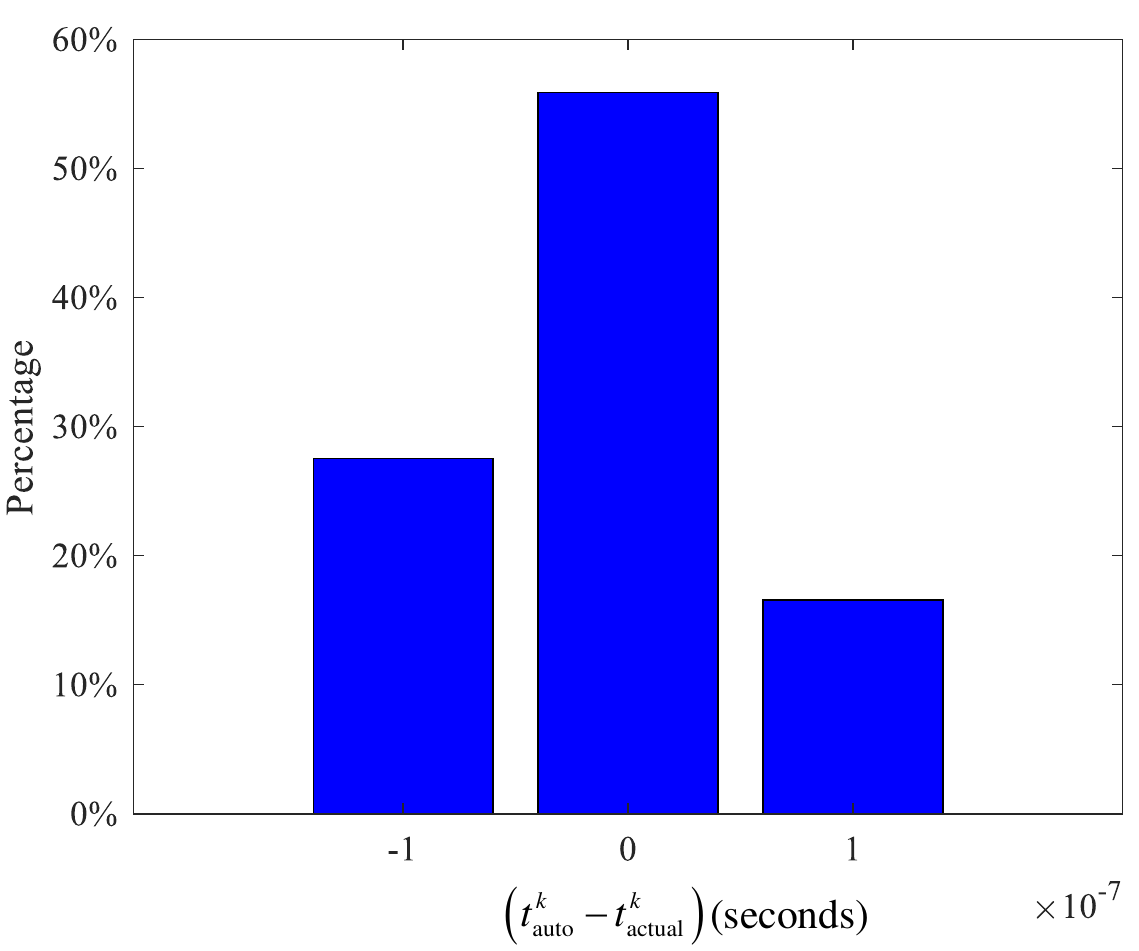}}
		\caption{The probability distribution of $\left( {t_{{\rm{auto}}}^k - t_{{\rm{actual}}}^k} \right)$.}
		\label{fig:auto_perf}
	\end{figure}
	
	We collected the statistics of $10^5$ UL packets in this experiment. The result is shown in Fig. \ref{fig:auto_perf}. From Fig. \ref{fig:auto_perf} we can see that $56\%$ of the UL packets' arrival times are perfectly provided by the Auto-trigger. Also, the absolute value of differences between $t_{{\rm{auto}}}^k$ and $t_{{\rm{actual}}}^k$ is always smaller than or equal to $1$ sample duration ($0.1\mu s$). Thus, we can conclude that the arrival times of UL packets provided by \autoTrig are highly accurate and are guaranteed to be $1$ sample offset at most. Considering that the UL packets' offsets among the users are smaller than $1$ sample duration, the safe range for doing the CP-cutting (no ISI after removing CP and doing FFT) is still wide enough. For an UL packet with a $16$-sample CP, the safe range is $14$-sample wide.
	
	\subsection{Performance of UL CFO Precoding} \label{subsec:exp_ul_prec_perf}
	In this subsection, we present the performance of UL precoding. We use the EVM as a metric to show the performance of the UL precoding for BPSK modulation. In an I-Q plane, the error vector is the distance vector between the ideal constellation point and the actual received constellation point. EVM is defined as follows:
	\begin{equation}
		EVM = \sqrt {\frac{{\frac{1}{{{r_{{\rm{recv}}}}}}\sum\nolimits_{i = 1}^{{r_{{\rm{recv}}}}} {{{\left| {{s_{{\rm{ideal}},i}} - {s_{{\rm{rec}},r}}} \right|}^2}} }}{{\frac{1}{{{r_{{\rm{unique}}}}}}\sum\nolimits_{i = 1}^{{r_{{\rm{unique}}}}} {{{\left| {{s_{{\rm{ideal}},i}}} \right|}^2}} }}},
	\end{equation}
	where ${s_{{\rm{ideal}},i}}$ is the ideal normalized constellation point for the $i$th received BPSK symbol, ${s_{{\rm{rec}},r}}$ is the normalized actual received BPSK symbol. ${r_{{\rm{unique}}}}$ is the number of unique symbols in the constellation (for BPSK, ${r_{{\rm{unique}}}} = 2$), and ${r_{{\rm{recv}}}}$ is the number of received symbols. Usually, ${r_{{\rm{recv}}}} \gg {r_{{\rm{unique}}}}$.
	
	The experiment was carried out on a ROFA system with one user. The AP broadcasts a DL packet in the DL phase. Then the user estimates the DL CFO using SLP CFO estimation with M-LTFs. After that, the user prepares a ROFA UL packet that is already known by the AP for the UL transmission. In this experiment, we fixed the user to use only 1 subcarrier for the UL transmission. Both UL packets with and without precoding transmit were tested.
	
	At the UL receiver side, the AP detects and demodulates the UL packet. We set the DL effective received SNR to be around $30dB$ by controlling the AP's transmit power, and varied the UL effective received SNR by controlling the user's transmit power. As a benchmark, we also measured the EVM in the UL transmissions in RTTS-SDR (OFDM-TDMA), using UL packets that carries the same amount of data.
	
	\begin{figure}
		\centering {\includegraphics[width=\oneFigWidth]{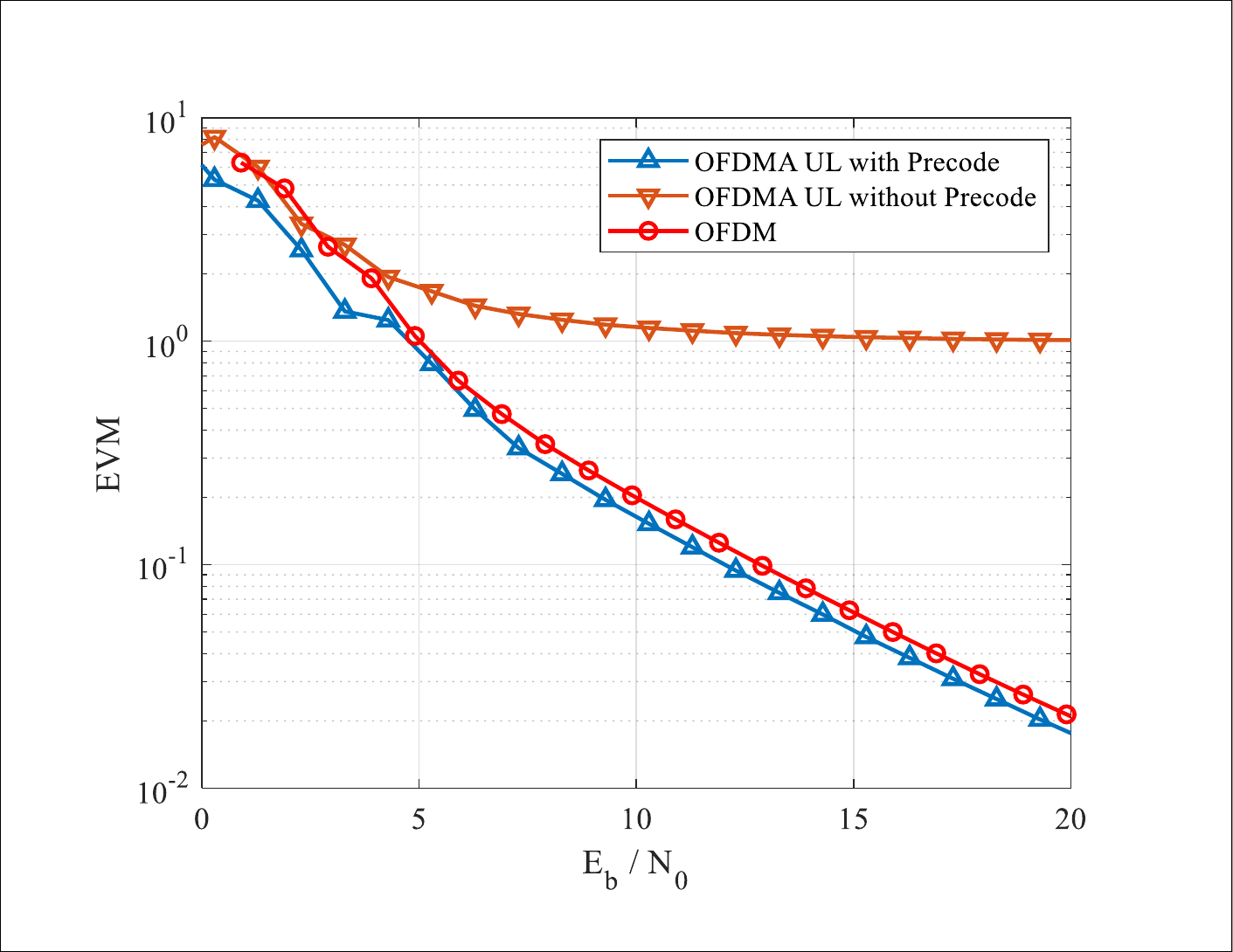}}
		\caption{The EVM of ROFA UL and OFDM systems.}
		\label{fig:evm_precode}
	\end{figure}
	
	We ran the experiment for $10^5$ rounds of UL and DL transmissions. Fig. \ref{fig:evm_precode} shows the EVM for BPSK modulation with and without the UL CFO precoding. Also, the EVM for BPSK modulation in OFDM transmission is plotted. For a fair comparison between the OFDMA and OFDM systems, we use ${E_b}/{N_0}$ for the x-axis to normalize the transmit power in different systems. As we can see, the EVM of ROFA UL has been reduced by a large amount with our precoding scheme compared with the scheme without precoding. ROFA UL even outperforms the OFDM system that uses pilots at the receiver to estimate and compensate for CFO. We believe that this is because OFDM uses the $4$ dedicated pilots in every OFDM symbol to compensate for the phase changes caused by the CFO, while the OFDMA UL does not: specifically, the power dedicated for pilots in OFDM is used for actual data transmission in the OFDMA system.
	
	\subsection{Reliability of ROFA versus RTTS-SDR} \label{subsec:exp_reliability}
	\subsubsection{Experiment Setup} \label{subsubsec:exp_setup}
	We carried out a series of experiments to compare the reliability of packet transmission in an OFDMA system and an OFDM-TDMA system (specifically, ROFA versus RTTS-SDR). In particular, we looked at the BER and the PER performance of the two systems. Since the DL transmission of both ROFA and RTTS-SDR use the same processing method (except for the subcarrier allocation), we focused on the UL transmission.
	
	To evaluate the BER and PER performances of ROFA UL, we performed experiments in the following two scenarios: (i) multiuser (MU) transmissions with balanced power; (ii) multiuser (MU) transmissions with imbalanced power. The detailed settings of these two scenarios are as follows:
	
	\textbf{\textit{Power-Balanced MU transmission.}} In this scenario, three users, ${X_1}$, ${X_2}$, and ${N_1}$, transmit UL packets to the AP, $C$, simultaneously using different subcarriers. The subcarrier allocations are as in TABLE \ref{tab:3_user_subcarrier_alloc}.
	\begin{table}
		\centering
		\captionsetup[subtable]{position = below}
		\captionsetup[table]{position=top}
		\caption{Subcarrier allocations for MU transmission.} \label{tab:subcarrier_alloc}
		\def\arraystretch{1.05}
		\begin{subtable}{0.4\linewidth}
			\centering
			\begin{tabular}{c|c}
				\hline
				\textbf{Users} & \textbf{Subcarriers} \\ \hline
				$X_1$  & $10, 13, 16$ \\ \hline
				$X_2$  & $11, 14, 17$ \\ \hline
				$N_1$  & $12, 15, 18$ \\ \hline
			\end{tabular}
			\caption{$3$ users}
			\label{tab:3_user_subcarrier_alloc}
		\end{subtable}%
		\hspace*{1em}
		\begin{subtable}{0.4\linewidth}
			\centering
			\begin{tabular}{c|c}
				\hline
				\textbf{Users} & \textbf{Subcarriers} \\ \hline
				$X_1$  & $10, 14, 18$ \\ \hline
				$X_2$  & $11, 15, 19$ \\ \hline
				$N_1$  & $12, 16, 20$ \\ \hline
				$N_2$   & $13, 17, 21$ \\ \hline
			\end{tabular}
			\caption{$4$ users}
			\label{tab:4_user_subcarrier_alloc}
		\end{subtable}
	\end{table}

	As per Section \ref{subsec:exp_setup}, users ${X_1}$, ${X_2}$, and ${N_1}$ are located in the same room and have similar distances to $C$. In order that all the users have approximately the same receive SNRs at $C$ (i.e., UL received SNRs), we calibrated\footnote{For an environment in which the channel is stable, the received SNR can be controlled by adjusting the transmit power of the USRPs.} the transmit power for each user. After that, we varied the transmit powers of all the users simultaneously to investigate the system performance under different SNRs. Details of the power adjustment can be found in Appendix \ref{append:tx_power_control}.
	
	The subcarrier allocations are set as in TABLE \ref{tab:3_user_subcarrier_alloc} to demonstrate the reliability performance of ROFA. Specifically, a subcarrier of a user is surrounded by the subcarriers of the other two users (except for the edge subcarriers $10$ and $18$). If the SLP CFO estimation and UL precoding are subpar, the ICI among the users may severely distort the receiving packets.
	
	\textbf{\textit{$3$-User Power Imbalanced MU transmission.}} Before the experiment, we calibrated ${X_1}$ and ${X_2}$ to have the same received SNRs at $C$. Then we varied the transmit power of ${N_1}$ to get the PER performance of ROFA UL in the power imbalanced scenario. Other settings of this experiment, including the subcarrier allocations, are the same as in the power-balanced case.
	
	This scenario is for the investigation of the near-far effect commonly seen in multiuser systems, including OFDMA systems. If the ICI between the radio nodes is not well compensated in the UL transmission, the near-far effect will degrade the performance of the users, especially that of the user with the weak transmit power.
	
	\textbf{\textit{4-User Power Imbalanced MU transmission with a NLoS Node.}} This scenario has four users transmit at the same time, ${X_1},{X_2},{N_1},{N_2}$. The settings are the same as the $3$-user case except for the subcarrier allocations. The subcarrier allocation plan is shown in TABLE \ref{tab:4_user_subcarrier_alloc}. ${X_1},{X_2}$ still has the same received SNRs at $C$. ${N_1}$ in this scenario, also has a fixed received SNR at $C$, which is stronger than that of ${X_1},{X_2}$. In addition, the new user, ${N_2}$, has varied received SNR at $C$ and was NLoS.
	
	This scenario was carried out to show that the NLoS user, ${N_2}$, still has good performance and does not affect the other users when there are strong users and weak user in the same network.
	
	\textbf{\textit{Benchmark: RTTS-SDR (OFDM-TDMA).}} For benchmarking, we ran the RTTS-SDR (an OFDM-TDMA system) on the same USRP platform. All the users take turns transmitting their packets and use up all the available subcarriers ($48$ out of $64$) in their transmissions. The transmit power of the users in RTTS-SDR are also calibrated so that they can have the same received SNR at $C$.
	
	Since RTTS-SDR is not a multiuser system that lets multiple users transmit their UL packets to the AP simultaneously, the power differences among the users do not affect the performance of the weak users in the system. Therefore, we only ran the system in a power-balanced setting, and compare its performance with power-balanced ROFA.
	
	\textbf{\textit{DL received SNR.}} Recall that the accuracy of the DL CFO estimation is affected by the DL received SNR. This accuracy will also impact the performance of the UL precoding. In the industrial environment, the AP typically has better amplifiers, better antennas, and larger power compared to the user devices.
	
	Since $C$ broadcasts the DL packets to all the users at the same time in the DL transmission, different users have different distances to $C$, we cannot make the DL received SNRs the same for all the users (adjusting the receiving gain does not affect the SNR). In our experiment, we ensured that the DL received SNR is at least $15dB$ at the farthest DL receiver $N_2$.
	
	\textbf{\textit{UL Received SNR.}} A conventional OFDM system uses the LTF in the preamble to estimate the signal power. However, the time-domain preambles of different users in ROFA UL are superimposed. 
	
	Since the LTFs of users are orthogonal in the frequency domain in ROFA UL, the AP can measure the average received signal power ${P_i}$ for user $i$, in the frequency domain. Specifically, when the AP processes the received LTF, it first converts the LTF to the frequency domain by an FFT, and then extracts the frequency domain LTF that belongs to user $i$ based on ${\cal S}_i$. By measuring the received signal power of each subcarrier in  ${\cal S}_i$, the AP can compute the average received signal power ${P_i}$. For the average noise power in each subcarrier $N_0$, the AP measures it during the Guard Interval before the arrival of a packet. The UL received SNR of user $i$ can then be computed by $\frac{{\left| {{\cal S}_i^{}} \right| \cdot {P_i}}}{{{N_{{\rm{FFT}}}} \cdot {N_0}}}$.
	
	For fair comparison, we also compute the UL received SNR for the OFDM-TDMA in the same way, in which ${\cal S}_i$ is the set of occupied tones.
	
	\subsubsection{Performance of Power-Balanced ROFA} \label{subsubsec:rofa_balance_perf}
\begin{figure}
		\centering {\includegraphics[width=\figWidth]{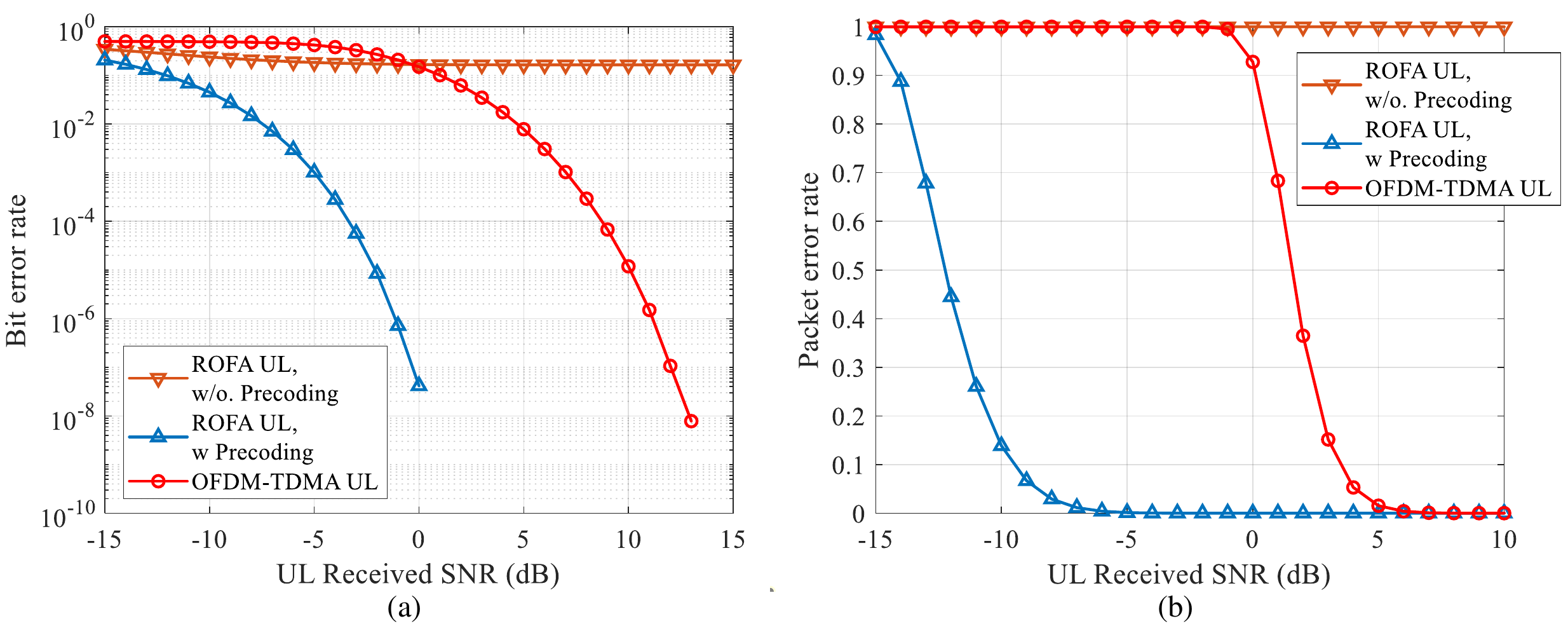}}
		\caption{The BER and PER performance of ROFA UL versus OFDM-TDMA UL.}
		\label{fig:ber_per_rofa_tdma}
	\end{figure}

	The BER and PER performance of the ROFA UL versus OFDM-TDMA UL is shown in Fig. \ref{fig:ber_per_rofa_tdma}, in which the UL received SNR is used as the x-axis.
	
	Fig. \ref{fig:ber_per_rofa_tdma}(a) presents the raw BER performance of two systems prior to channel decoding. To achieve the same BER, OFDM-TDMA UL needs around $12dB$ more of the received SNR.  Users in OFDM-TDMA need to spend $12dB$ more on their transmit powers for the same BER performance.
	
	Fig. \ref{fig:ber_per_rofa_tdma}(b) shows the PER after channel decoding. The AP used soft-bit information for channel decoding. From the figure we can see that, with channel coding, the ROFA UL still outperforms OFDM-TDMA UL. For a given PER target, ROFA can also be up to $12dB$ more power-efficient for the same performance.
	
	We remark that ROFA users only use a few subcarriers while the noise power is calculated over the whole band. Thus, the performance of ROFA users is still acceptable when the SNR is in the negative range. For example, a $3$-subcarrier user that has an average received SNR $\frac{P_i}{N_0} = 2dB$ on its subcarriers has the UL received SNR $\frac{{\left| {{\cal S}_i^{}} \right| \cdot {P_i}}}{{{N_{{\rm{FFT}}}} \cdot {N_0}}}$ equal to $2 + 10{\log _{10}}\left( {3/64} \right)(dB) =  - {\rm{11}}{\rm{.29dB}}$. In other words, a ROFA user is concentrating its transmit power on $3$ subcarriers as opposed to $64$ subcarriers, while an OFDM-TDMA user is spreading the same transmit power over all the subcarriers. 
	
	\subsubsection{Performance of Power Imbalanced ROFA} \label{subsubsec:rofa_imbalance_perf}
\begin{figure}
		\centering {\includegraphics[width=\figWidth]{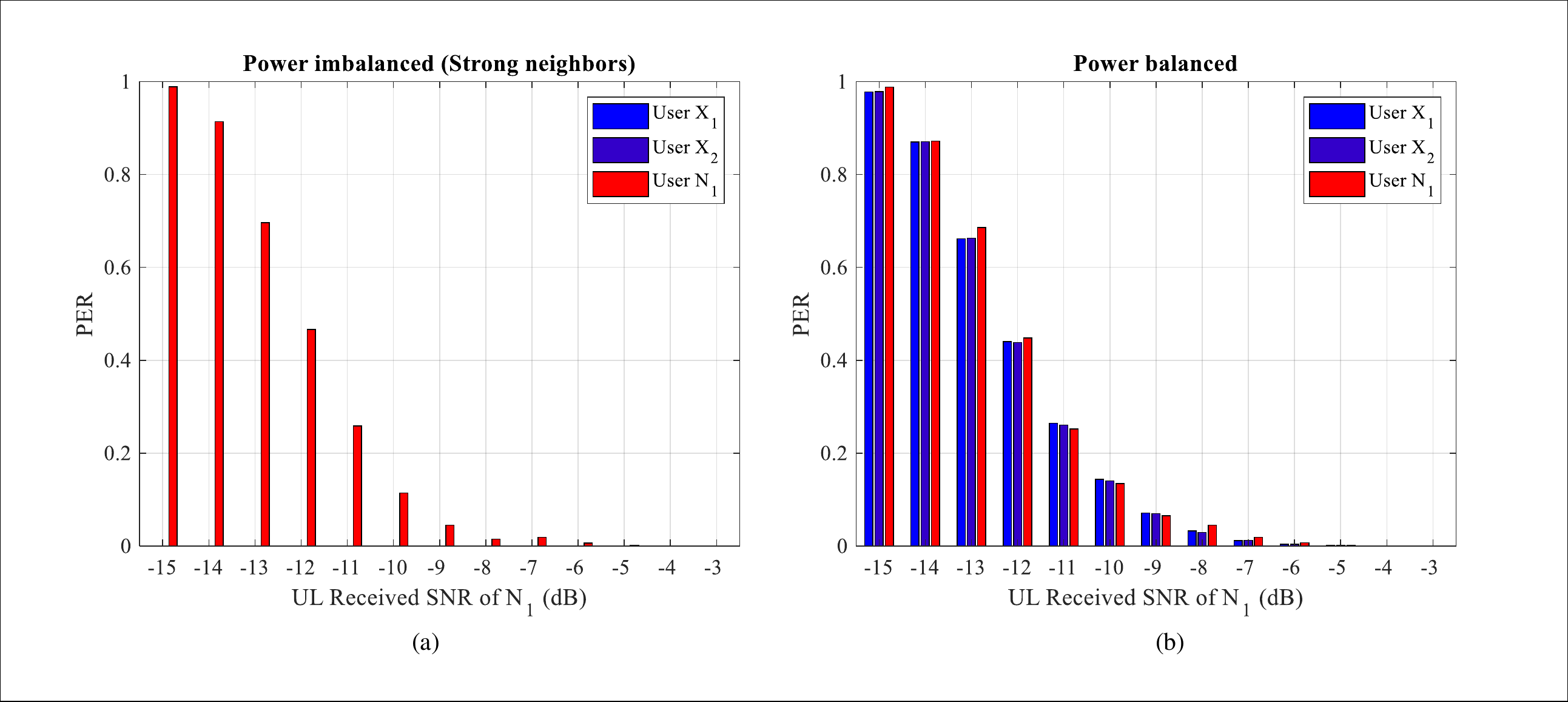}}
		\caption{The PER performance of three users in (a) Power imbalanced and (b) Power-balanced scenarios. The UL received SNRs of ${X_1}$ and ${X_2}$ in (a) are both $0dB$. In (b), three users have the same UL received SNRs.}
		\label{fig:per_imbal}
	\end{figure}

	We next compare the PER performance of the ROFA UL when the power is balanced or imbalanced, as shown in Fig. \ref{fig:per_imbal}. In this figure, different bars represent the PER of different users.
	
	In Fig. \ref{fig:per_imbal}(a), we show the case when two users have strong power and one user has weak power. Specifically, we let both ${X_1}$ and ${X_2}$ (strong users) transmit UL packets with UL Received SNRs equal to $0dB$. Then we varied the transmit gain of ${N_1}$ (weak user) to vary its UL received SNR from $-15dB$ to $-3dB$. We can only see the red bars because ${X_1}$ and ${X_2}$ have enough SNR to achieve zero PER. We also plot the PER of each user in power-balanced ROFA for benchmarking purposes in Fig. \ref{fig:per_imbal}(b). Note that in this benchmarking power-balanced ROFA, all the users increase their transmit gains and the UL received SNRs vary from $-15dB$ to $-3dB$. Comparing the red bars in Fig. \ref{fig:per_imbal}(a) and Fig. \ref{fig:per_imbal}(b), we can see that $N_1$ has similar performance in both power-balanced and power imbalanced scenarios, regardless of the other users' transmit power.
	
	The result indicates that ROFA UL is robust against the near-far effect, as the user with weak transmit power (i.e., $\text{UL received SNR}<-10dB$) will not have performance degradation even if its neighbor users have strong transmit powers.
	
\begin{figure}
		\centering {\includegraphics[width=\figWidth]{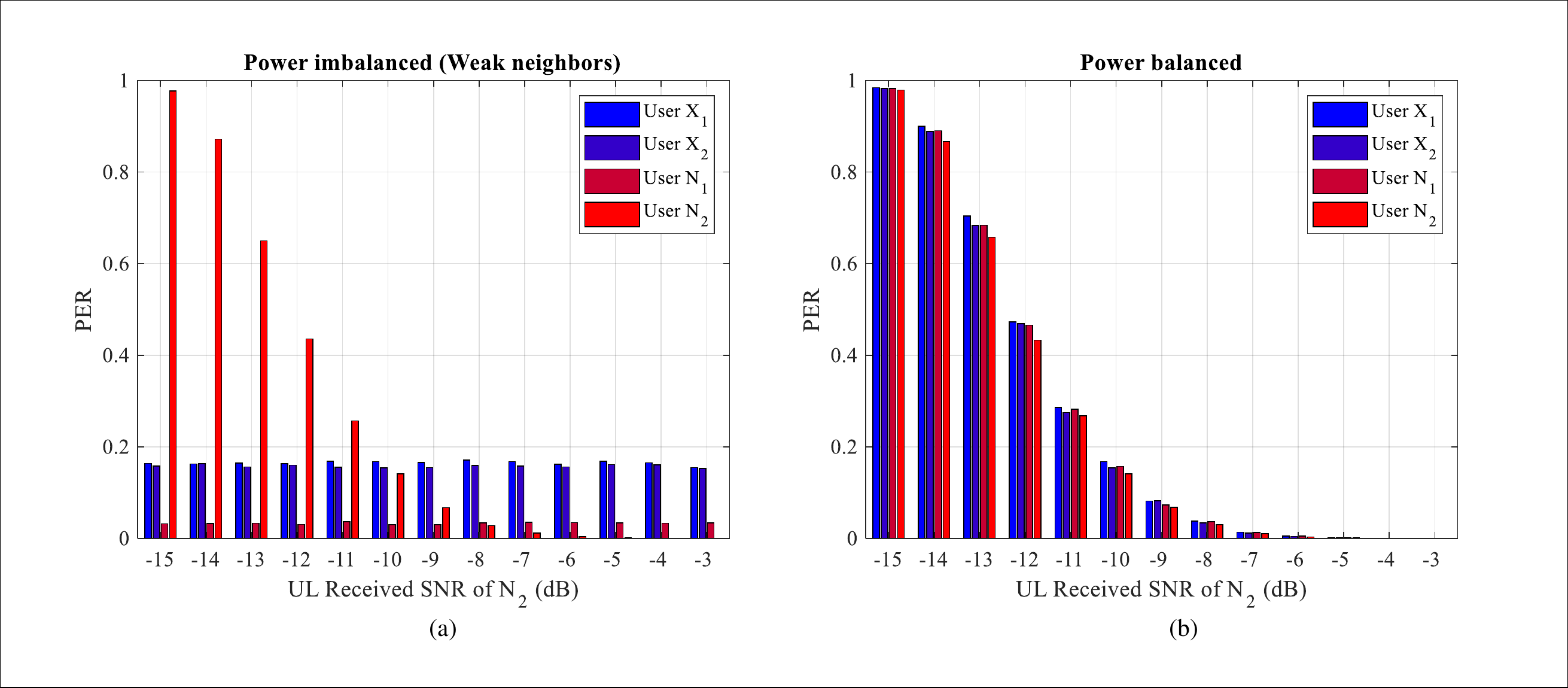}}
		\caption{The PER performance of (a) power imbalanced, and (b) power-balanced, when the system has four users.}
		\label{fig:per_imbal_bal_weak}
	\end{figure}

	We also verified the above conclusion by making $N_1$ the strong user that has UL received SNR equal to $-8dB$, and ${X_1}$, ${X_2}$ the medium users that has UL received SNRs equal to $-10dB$. We varied the transmit power of $N_2$. The result in Fig. \ref{fig:per_imbal_bal_weak}(a) shows that when $N_2$ (a NLoS user) increases its transmit gain to cause the UL received SNR to vary from $-15dB$ to $-3dB$, the PER performances of all the other users are not affected. Furthermore, Fig. \ref{fig:per_imbal_bal_weak}(b) shows that $N_2$ has a similar performance in this four-user system if all the users have the same UL received SNRs. We can conclude that a user (either LoS or NLoS) has the same performance regardless of other users' transmit power in the same network.
	
	\subsubsection{With Low-cost oscillators} \label{subsubsec:low_cost_osc_perf}
	The oscillators in the USRPs are relatively good. We followed the method in \cite{liang2020design_jiot} to emulate the effects of the low-cost oscillators in our USRP experiments. Specifically, we increased both the UL CFO and the DL CFO by $20$ times (i.e., relative CFO is equal to $1000Hz \times 20 = 20kHz$) by artificially introducing CFO to the raw transmit signal. We performed the experiments for the low-cost oscillator in the aforementioned two scenarios.
	
	\begin{figure}
		\centering {\includegraphics[width=\figWidth]{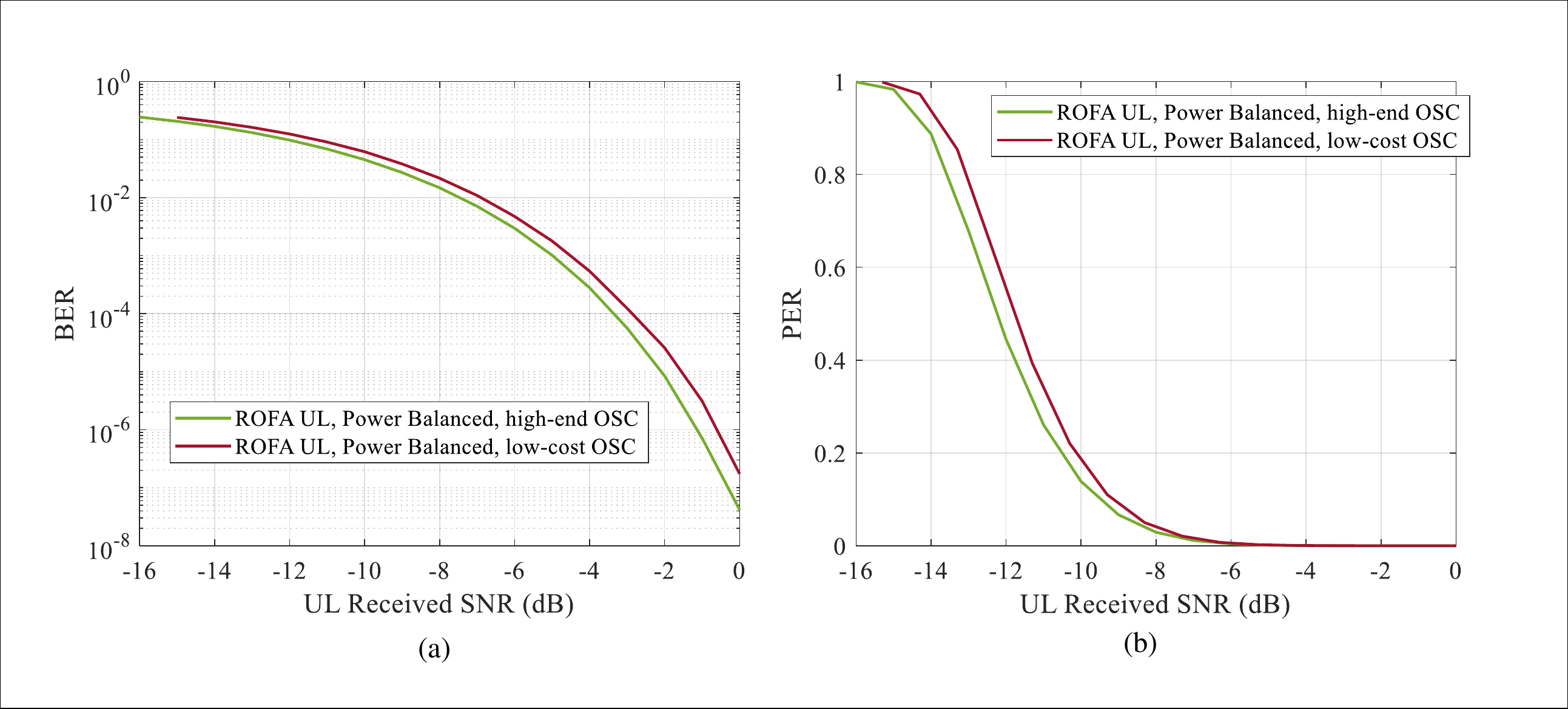}}
		\caption{The (a) BER and (b) PER performances when emulating the low-cost oscillators.}
		\label{fig:low_cost_perf}
	\end{figure}

	Fig. \ref{fig:low_cost_perf} shows that power-balanced ROFA UL maintains similar performance even when a low-cost oscillator is used. Specifically, Fig. \ref{fig:low_cost_perf}(a) shows that the low-cost oscillators degrade the BER performance of ROFA UL by only $0.7dB$. We can also see the same amount of degradation in the PER performance when channel coding is used, as shown in Fig. \ref{fig:low_cost_perf}(b). Similar performance is also observed in power-imbalanced ROFA UL.
	
	Note that the low-cost oscillator also causes symbol misalignment among the users in the UL transmission. But we can count on the time-synchronization mechanism which is explained in Section \ref{subsec:system_design_ul_sync}, to fulfill the synchronization requirement. Since the time-synchronization of the users is executed in every DL packet, the time synchronization will not be a major problem.
	
	\subsection{Ultra-reliability for Short Packets} \label{subsec:exp_ultra_relia}
	Packets that have ultra-reliability requirements (e.g., sensor data and control commands) typically need to transmit very little data. We investigated the use of short packets of $12$ bytes.
	
	In this experiment, ROFA UL uses BPSK modulation, rate $1/2$ convolutional code, and $3$ subcarriers for each user, yielding an overall payload of $64$ OFDM symbols. For benchmarking, we also conducted the packet transmission with the same number of bytes in (i) OFDM-TDMA ($48$ subcarriers) and (ii) a "broader" ROFA in which each user is allocated $13$ subcarriers instead of $3$ subcarriers. With $13$ subcarriers, each user in the "broader" ROFA requires the same bandwidth ($10MHz \times \frac{{13}}{{64}} = 2.0313MHz$) as the smallest RU in 802.11ax\footnote{ In 802.11ax, the smallest RU is $26$ subcarriers when operating in $20MHz$. The bandwidth for this RU is $2.0313MHz$.}. The numbers of OFDM symbols for the payload in OFDM-TDMA and "broader" ROFA are $4$ and $15$, respectively.
	

	\begin{figure}
		\centering {\includegraphics[width=\oneFigWidth]{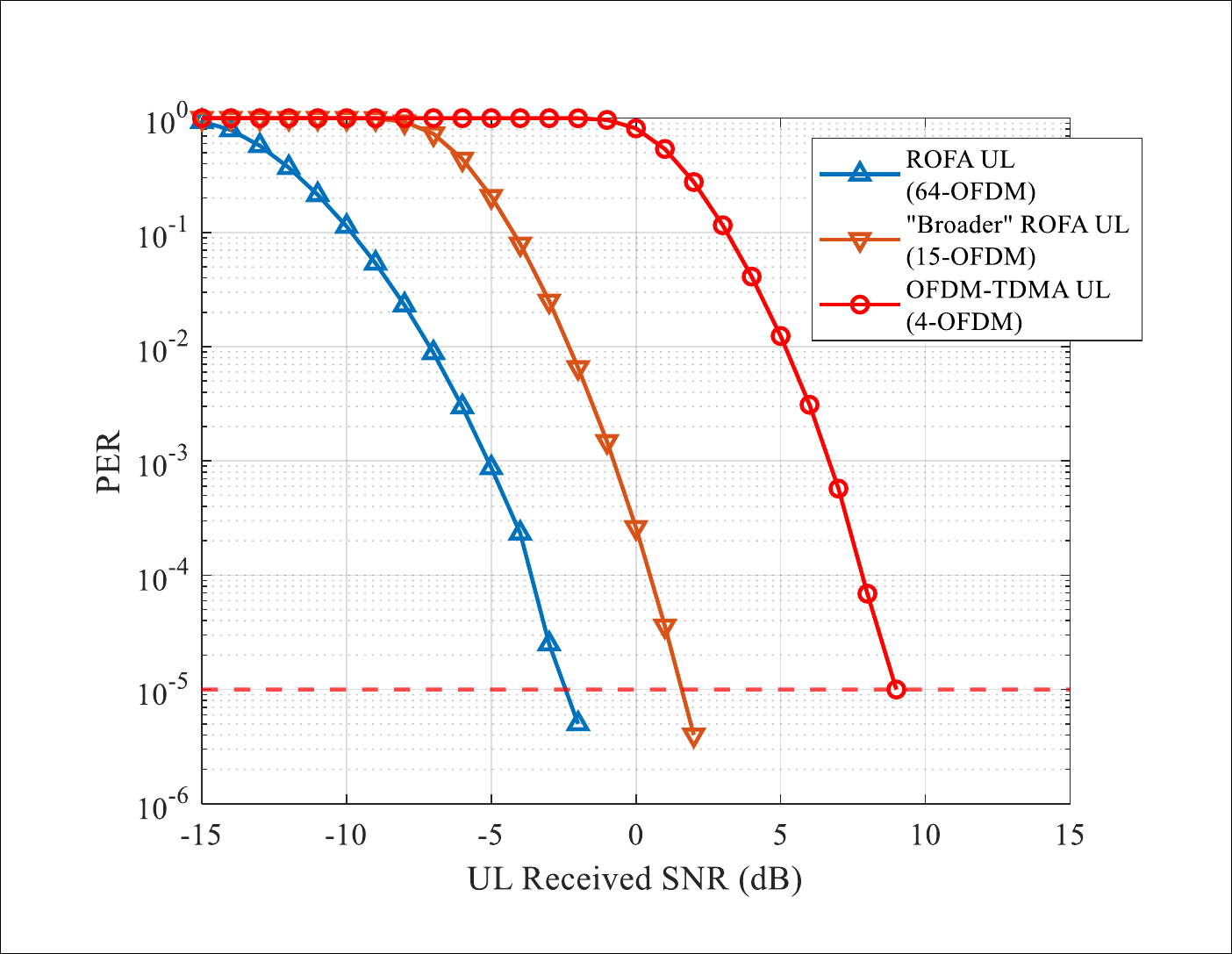}}
		\caption{The PER versus UL received SNR when three systems transmit short packets.}
		\label{fig:ultra_reliable}
	\end{figure}

	In this experiment, each of the $3$ users transmits ${10^6}$ packets. The results are shown in Fig. \ref{fig:ultra_reliable}. We can see from the figure that ROFA UL can deliver a short packet successfully with $99.999$-percentile certainty (i.e., $PER < {10^{ - 5}}$) when the UL received SNR is around $-2.5dB$. ROFA UL achieves the same reliability as ``Broader'' ROFA UL and OFDM-TDMA UL with $4dB$ less power and $11.5dB$ less power, respectively. For the same $12$ bytes of data, since ROFA UL uses the fewest subcarriers, it has the longest packet duration compared with ``Broader'' ROFA UL and OFDM-TDMA UL. Although the difference in packet durations causes some degradations compared with the results in \ref{fig:ber_per_rofa_tdma}(b) (the PER gap between ROFA UL and OFDM-TDMA UL decreases from $12dB$ to $11.5dB$), ROFA still provides the most reliable packet transmission among the three.
	
	\section{Conclusion} \label{sec:conclusion}
	\textcolor{\txtColor}{We have presented and experimentally evaluated a system, ROFA, that supports ultra-reliable packet delivery in an infrastructure network through OFDMA. OFDMA, especially OFDMA UL, requires the transmitters to align their UL packets to within-CP. Furthermore, the UL receiver needs to compensate for the CFOs among the UL signals from multiple transmitters. It is challenging to design a mechanism for such system when each radio node is driven by an independent clock. ROFA, as a system targeting for URC, tackles the challenges with the following designs: (i) a downlink-coordinated time-synchronization mechanism that synchronizes the UL transmissions of users, with at most $0.1\mu \sec$ timing offset; (ii) an UL packet reception synchronization method, \textit{Auto-trigger}, that does away with the need to detect packet arrival before packet decoding, hence eliminating the possibility of packet misdetection that may compromise system reliability in the low-SNR regime; (iii) an UL precoding mechanism that reduces the CFOs between users and the AP to a negligible level, by making use of \textit{CFO reciprocity} and \textit{SLP CFO estimation}, hence largely increasing the packet decoding probability.}
	
	Extensive experiments have been carried out on the real-time implementation of ROFA. The results indicate that ROFA can achieve the same BER and PER with $12dB$ less transmit power compared to OFDM-TDMA when each user in ROFA uses three of the available subcarriers only. The ICI and near-far effect commonly seen in OFDMA systems are negligible in ROFA, thanks to the accuracy of the CFO precoding at the transmitter. For ultra-reliable communication ($\rm{PER}<10^{-5}$), ROFA outperforms OFDM-TDMA and ``broader ROFA'' (which has the same bandwidth as 802.11ax's smallest RU) by $11.5dB$ and $4dB$, respectively. 
	
	The OFDMA scheme of 802.11ax aims for added flexibility in spectrum usage and potential  higher throughput. There is much overhead, however, to ensure backward compatibility with earlier 802.11 standards. ROFA, as a clean-slate design, allows users to use fewer subcarriers, removes potential packet misdetection with the use of STF with a new \autoTrig mechanism, and precodes packets to remove ICIs among different users. ROFA is designed with one end target: providing ultra-reliable packet transmission for industrial applications. As such, unlike 802.11ax which is intended for all sorts of different applications, ROFA can better focus on the needs of the niche application it is intended for, without the unnecessary overhead and backward-compatibility mechanisms in 802.11ax.

	
	%

	\appendices
	\section{Analysis of SLP CFO estimation's residual CFO} \label{append_analy_residual_cfo}
	Recall that the CFO estimation using correlation cannot measure a CFO in a period if the phase rotation is more than $2 \pi$ in this period. Otherwise, the erroneous estimated CFO value will leave a  more-than $2 \pi$ residual CFO in the signals after the compensation. 
	
	The above statement also applies to the SLP CFO estimation. Suppose the actual phase rotation between LTF and P-LTF, after Fine CFO compensation, is ${\theta _{{\rm{true}}}}$. The SNR is assumed fixed. Let $\Delta \mathord{\buildrel{\lower3pt\hbox{$\scriptscriptstyle\frown$}} 
		\over f} $ denote the CFO estimated by SLP CFO estimation, we know that $\Delta \mathord{\buildrel{\lower3pt\hbox{$\scriptscriptstyle\frown$}} 
		\over f} {\lambda _{\rm{P}}} \in [0,2\pi ]$, ${\lambda _{\rm{P}}}$ is the distance between LTF and P-LTF. The CFO per sample in between the LTF and the P-LTF is $\Delta {\theta _{{\rm{true}}}} = {\theta _{{\rm{true}}}}/{\lambda _{\rm{P}}}$. The relationship between $\Delta {\theta _{{\rm{true}}}}$ and $\Delta \mathord{\buildrel{\lower3pt\hbox{$\scriptscriptstyle\frown$}} 
		\over f} $ can be written as
	\begin{equation}
		{\theta _{{\rm{true}}}} = \Delta {\theta _{{\rm{true}}}}{\lambda _{\rm{P}}} = \left( {\Delta \mathord{\buildrel{\lower3pt\hbox{$\scriptscriptstyle\frown$}} 
				\over f}  + \varepsilon } \right) \cdot {\lambda _{\rm{P}}} + m \cdot 2\pi , 
	\end{equation}
	where $m$ is an integer, and $\varepsilon $ is the estimation error of the SLP CFO estimation.
	
	When ${\theta _{{\rm{true}}}} \in \left[ {0,2\pi } \right]$, we have $m = 0$ and $\Delta {\theta _{{\rm{true}}}} = \Delta \mathord{\buildrel{\lower3pt\hbox{$\scriptscriptstyle\frown$}} 
		\over f}  + \varepsilon $. If $\Delta \mathord{\buildrel{\lower3pt\hbox{$\scriptscriptstyle\frown$}} 
		\over f} $ is used to compensate for the CFO, the residual CFO after compensation is equal to $\varepsilon $. When ${\theta _{{\rm{true}}}} \in \left[ {2\pi ,4\pi } \right]$, we know that $m = 1$, hence, $\Delta {\theta _{{\rm{true}}}} = \Delta \mathord{\buildrel{\lower3pt\hbox{$\scriptscriptstyle\frown$}} 
		\over f}  + \varepsilon  + \frac{{2\pi }}{{{\lambda _{\rm{P}}}}}$. By compensating for the CFO with $\Delta \mathord{\buildrel{\lower3pt\hbox{$\scriptscriptstyle\frown$}} 
		\over f} $, the residual CFO is equal to $\varepsilon  + \frac{{2\pi }}{{{\lambda _{\rm{P}}}}}$. In general, when ${\theta _{{\rm{true}}}} \in \left[ {m \cdot 2\pi ,(m + 1) \cdot 2\pi } \right]$, the residual CFO after compensation is $\varepsilon  + \frac{{m \cdot 2\pi }}{{{\lambda _{\rm{P}}}}}$. Note that the unit in $\varepsilon  + \frac{{m \cdot 2\pi }}{{{\lambda _{\rm{P}}}}}$ is rad/sample. In general, with good estimation $\varepsilon $ is relatively small $(\varepsilon  \to 0)$, so that we would expect to see the CDF after compensation to have jumps at $\frac{{m \cdot 2\pi }}{{{\lambda _{\rm{P}}}}}, m=\{1, 2, … \}$, and remain flat in between two jumps. 
	
	\section{Transmit Power Control in USRP} \label{append:tx_power_control}
	There are two ways to control the transmit power on the USRP platform: control the transmit samples' amplitude; control the transmission gain of the amplifier on the USRP daughterboard. In our experiments, we keep the transmit samples' amplitude unchanged and varied the transmission gain to control the transmit power. Specifically, for both ROFA and RTTS-SDR, we normalized the amplitudes of the transmit samples at the transmitter, so that the sample amplitude does not affect the transmit power.
	
	In power-balanced MU transmission, the transmit gains of the different users were calibrated to have approximately the same received SNRs at $C$. After that, we varied the transmission gains for all the users simultaneously, so that their received SNRs are always approximately the same.
	
	In power imbalanced MU transmission, we calibrated the transmission gains of ${X_1}$ and ${X_2}$. We then varied the transmission gain of ${N_1}$ to change its transmit power for the $3$-user scenario. For the $4$-user scenario, we fixed the transmission gain of $N_1$, and changed the transmission gain of $N_2$.

%
%

	\ifCLASSOPTIONcaptionsoff
	\newpage
	\fi

	
	
	%

	\bibliographystyle{IEEEtran}
	\bibliography{References}
	
	%
	
	
	
	
	
	
	

\end{document}